%% file: eosmu.tex
\documentclass[11pt,a4paper]{article}
\pdfoutput=1
\usepackage{jheppub}
\usepackage{amsfonts,amssymb,epsfig,verbatim,mathbbol,amstext,amsmath,psfrag}
\usepackage{graphicx}
\usepackage[latin2]{inputenc}
\usepackage{t1enc}
\usepackage{amsmath}
\usepackage{subfigure}
\usepackage{dsfont}
\usepackage{slashed}
\usepackage{multirow}
\usepackage{lscape}
\usepackage{rotating}
\usepackage{wrapfig}
\usepackage{color}
\usepackage{array}

\usepackage{setspace}

\usepackage{colordvi}
\usepackage{lscape}
\usepackage{rotfloat}
\usepackage{amssymb}

\usepackage{cmap}

\newcommand{\dd}{\textmd{d}}
\newcommand{\be}{\begin{equation}}
\newcommand{\ee}{\end{equation}}

\newcommand{\Z}{\mathcal{Z}}

\long\def\symbolfootnote[#1]#2{\begingroup%
\def\thefootnote{\fnsymbol{footnote}}\footnote[#1]{#2}\endgroup} 

\newcommand{\Wuppertal}{Bergische Universit\"at Wuppertal, Theoretical Physics, 42119 Wuppertal, Germany.}
\newcommand{\Budapest}{E\"otv\"os University, Theoretical Physics, P\'azm\'any P. S 1/A, H-1117, Budapest, Hungary.}
\newcommand{\Regensburg}{Institute for Theoretical Physics, Universit\"at Regensburg, D-93040 Regensburg, Germany.}
\newcommand{\Juelich}{J\"ulich Supercomputing Centre, Forschungszentrum J\"ulich, D-52425 J\"ulich, Germany.}
\newcommand{\Torino}{Dipartimento di Fisica, Universita` degli Studi di Torino and INFN, Sezione di Torino, I-10125 Torino, Italy.}

\begin{document}

\preprint{WUB/12-10}

\title{QCD equation of state at nonzero chemical potential: continuum results with physical quark masses at order $\mu^2$}

\author[1]{Sz.~Bors\'anyi,}
\author[2]{G.~Endr\H{o}di,}
\author[1,3,4]{Z.~Fodor,}
\author[3]{S.~D.~Katz,}
\author[1,4]{S.~Krieg,}
\author[5]{C.~Ratti,}
\author[1]{K.~K.~Szab\'o}
\affiliation[1]{\Wuppertal}
\affiliation[2]{\Regensburg}
\affiliation[3]{\Budapest}
\affiliation[4]{\Juelich}
\affiliation[5]{\Torino}

\emailAdd{borsanyi@uni-wuppertal.de}
\emailAdd{gergely.endrodi@physik.uni-regensburg.de}
\emailAdd{fodor@physik.uni-wuppertal.de}
\emailAdd{katz@bodri.elte.hu}
\emailAdd{s.krieg@fz-juelich.de}
\emailAdd{ratti@to.infn.it}
\emailAdd{szaboka@general.elte.hu}

\abstract{
We determine the equation of state of QCD for nonzero chemical potentials via a Taylor expansion of the pressure. The results are obtained for $N_f=2+1$ flavors of quarks with physical masses, on various lattice spacings. We present results for the pressure, interaction measure, energy density, entropy density, and the speed of sound for small chemical potentials. At low temperatures we compare our results with the Hadron Resonance Gas model. We also express our observables along trajectories of constant entropy over particle number. A simple parameterization is given (the Matlab/Octave script {\tt parameterization.m}, submitted to the arXiv along with the paper), which can be used to reconstruct the observables as functions of $T$ and $\mu$, or as functions of $T$ and $S/N$.
}


\maketitle

\section{Introduction}

Quantum Chromodynamics (QCD) is the theory of the strong interactions. It predicts that at high temperatures and/or chemical potentials strongly interacting matter exhibits a transition which separates the hadronic, confined phase and the quark-gluon plasma (QGP) phase. This transition took place during the evolution of the early universe, and can also be reproduced in contemporary heavy-ion collision experiments. 
A remarkable outcome of these experiments is that the hot, strongly interacting matter behaves much like a nearly ideal relativistic fluid, and thus can be described by relativistic hydrodynamics, see, e.g. Refs.~\cite{Teaney:2000cw,Teaney:2001av,Kolb:2003dz,Jacobs:2004qv,Romatschke:2009im}. A crucial aspect of such a description is the relation between local thermodynamic quantities, i.e. the equation of state (EoS). 
The parameters of the EoS are the temperature $T$ and the chemical potential $\mu$. The phase structure of QCD as a function of these parameters constitutes the phase diagram on the $T-\mu$ plane.
Since ideal relativistic hydrodynamics is entropy conserving -- if the ideal fluid description is indeed appropriate -- the ratio of the entropy and the particle number $S/N$ is expected to remain constant during the expansion of the QGP in a heavy ion collision. For the hydrodynamic description of the plasma, the equation of state along such an isentropic line is therefore particularly important~\cite{Hung:1997du,Toneev:2000ym,Bluhm:2007nu}.

The most powerful tool to study the EoS and the phase diagram is lattice QCD. At zero chemical potential the transition was determined to be an analytic crossover~\cite{Aoki:2006we} and the corresponding pseudocritical temperatures $T_c$ were calculated. Although initially inconsistent~\cite{Cheng:2006qk,Aoki:2006br}, studies using different fermionic discretizations~\cite{Cheng:2006qk,Bazavov:2011nk} and~\cite{Aoki:2006br,Aoki:2009sc,Borsanyi:2010bp} now converge to the same continuum limit~\cite{Aoki:2006br,Aoki:2009sc,Borsanyi:2010bp,Bazavov:2010sb,Bazavov:2011nk}. A variety of studies about the EoS can also be found in the literature, for the most recent developments see e.g. Refs.~\cite{Ejiri:2005uv,Bernard:2006nj,Cheng:2009zi,Borsanyi:2010cj}. 
Here different fermionic discretizations still give significantly different results for the trace anomaly.
This discrepancy was discussed recently at this year's lattice conference, and the reason for it was suggested~\cite{LombardoPlenary} to lie in our continuum extrapolation scheme. 
We emphasize that this is not the case, as our continuum extrapolation gives within errors the same limit, independently of whether we employ tree-level improvement in the trace anomaly or not, see Fig. 8 of Ref.~\cite{Borsanyi:2010cj} (for a few temperatures) and the preliminary results in Fig. 1 of Ref.~\cite{Borsanyi:2012vn} (for the whole temperature range).
Since the results for different discretizations should agree in the continuum limit,
one expects that similarly to the case of $T_c$ the findings will converge
for the equation of state, too. This is certainly a task for the near future.

While several lattice developments about the $\mu=0$ EoS are present, at $\mu>0$ lattice simulations are hindered by the sign problem, making standard Monte-Carlo methods based on importance sampling impossible. Starting with Ref.~\cite{Fodor:2001au} a renewed interest has been seen for $\mu>0$ lattice questions. Since then several alternatives were developed to circumvent the sign problem which include reweighting, Taylor-expansion in $\mu$, analytic continuation from imaginary $\mu$, the density of states method, or using the canonical ensemble; each of them have their pros and cons. For a recent review about the subject see, e.g. Ref.~\cite{Fodor:2009ax} and references there. 
Most studies of the finite $\mu$ EoS were performed using reweighting~\cite{Csikor:2004ik}, Taylor-expansion~\cite{Fodor:2002km,Allton:2003vx,Allton:2005gk,Bernard:2007nm,Basak:2009uv,DeTar:2010xm} or recently, in the imaginary chemical potential formalism~\cite{Takaishi:2010kc}. 
In this paper we follow the second approach (which is the truncation of the first one, see Ref.~\cite{Fodor:2009ax}) and carry out the leading order Taylor-expansion of thermodynamic observables. In this approach one is naturally constrained to small chemical potentials, however, the reliability of the results can be tested by comparing with the next-to-leading order terms. For the Taylor method, unlike for analytic reweighting, present computational resources allow one to extrapolate lattice data to the continuum limit and thereby control the main source of systematic error of lattice calculations.

Using thermodynamic observables at $\mu=0$ and the Taylor-coefficients of the pressure we calculate the equation of state of $2+1$ flavor QCD for small chemical potentials. We also search for trajectories of constant $S/N$ on the phase diagram to determine the EoS along the isentropic lines.
We use physical quark masses which is a crucial ingredient since most thermodynamic observables depend strongly on the quark masses. Lattice results are obtained in the temperature range $125 \textmd{ MeV} <T< 400 \textmd{ MeV}$. At low temperatures we compare the results to the Hadron Resonance Gas (HRG) model prediction. We define a simple function which connects the HRG and the lattice data, and can be used to describe thermodynamic quantities for the whole range $0<T<400 \textmd{ MeV}$ and small chemical potentials.

\section{Equation of state at nonzero \boldmath \texorpdfstring{$\mu$}{mu}}
\label{sec:formulation}

All information about the properties of a thermodynamic system is encoded in the grand canonical partition function $\Z$. For QCD, the parameters of $\Z$ are the quark masses $m_i$, the chemical potentials $\mu_i$ and the gauge coupling $g^2$, which, through the $\beta$-function, determines the temperature $T$. Here $i$ runs over the quark flavors, $i=u,d,s$. We consider equal masses for the up and down flavors, $m_{ud}\equiv m_u=m_d$. Moreover, we fix the quark masses $m_{ud}$ and $m_s$ to their physical value (see section~\ref{sec:simdet}). The remaining free parameters are thus the temperature and the chemical potentials. 

The primary observable for the study of the equation of state is the pressure, which, in the thermodynamic limit can be written as
\be
p(T,\{\mu_i\}) = \frac{T}{V}\log\Z(T,\{\mu_i\}),
\label{eq:pdef}
\ee
where $V$ denotes the three-dimensional volume of the system.
From the pressure the (net) quark number densities $n_j$ and the energy density $\epsilon$ are derived as
\be
\begin{split}
n_j(T,\{\mu_i\}) &= \frac{T}{V} \frac{\partial \log\Z}{\partial \mu_j},\\
\epsilon(T,\{\mu_i\}) &= \frac{T^2}{V} \frac{\partial \log\Z}{\partial T} + \sum_j \mu_j n_j(T,\{\mu_i\}).
\end{split}
\label{eq:edef}
\ee
Time reversal symmetry ensures that the partition function is an even function of the chemical potentials. The leading order Taylor-expansion of the dimensionless pressure is thus
\be
\frac{p(T,\{\mu_i\})}{T^4} = \frac{p(T,\{0\})}{T^4} + \frac{1}{2} \sum_{i,j} \frac{\mu_i\mu_j}{T^2} \chi_2^{ij},
\label{eq:pmu}
\ee
with
\be
\chi_2^{ij} \equiv \frac{T}{V} \frac{1}{T^2}\left.\frac{\partial ^2 \log\Z}{\partial \mu_i \partial \mu_j}\right|_{\mu_i=\mu_j=0}.
\label{eq:chi2def}
\ee

We proceed by prescribing the relation between the chemical potentials for the various flavors. A possible choice is the full baryonic chemical potential,
\be
\mu_B/3 \equiv \mu_u=\mu_d=\mu_s,
\label{eq:muB}
\ee
which excites strangeless as well as strange baryons. However, to describe the relevant situation for a heavy ion collision, where the net strangeness density is approximately zero throughout the experiment, it is advantageous to consider a strangeless baryonic chemical potential. 
Besides setting $\mu_u=\mu_d$, one can tune the strange chemical potential such that $n_s=0$ is fulfilled. Taking the derivative of the expansion of the pressure, Eq.~(\ref{eq:pmu}), with respect to $\mu_s$ gives the net strangeness density. The constraint $n_s=0$ then leads to a temperature-dependent relation between $\mu_s$ and $\mu_u$, given in terms of the fluctuations $\chi_2^{us}$ and $\chi_2^{ss}$. Putting these together, we define the `light' baryonic chemical potential $\mu_L$ as
\be
\mu_{L}/3\equiv \mu_u=\mu_d,\quad\quad 
\mu_s = -2\frac{\chi_2^{us}}{\chi_2^{ss}} \mu_u,
\label{eq:muL}
\ee
and in the following $\mu$ will denote either $\mu_L$ or $\mu_B$. In the same manner, the particle number density will denote either the `light' density $n=n_L$ or the baryonic density $n=n_B$.

Another quantity of interest is the trace anomaly,
\be
I(T,\mu)\equiv \epsilon(T,\mu)-3p(T,\mu),
\ee
which is calculated using Eqs.~(\ref{eq:pdef}--\ref{eq:chi2def}) as
\be
\frac{I(T,\mu)}{T^4} = T \frac{\partial }{\partial T} \frac{p(T,\mu)}{T^4} \,+\, \frac{\mu^2}{T^2}\chi_2 =
\frac{I(T,0)}{T^4} \,+\, \frac{\mu^2}{2T} \frac{\partial \chi_2}{\partial T}.
\label{eq:imu}
\ee
At $\mu=0$ the inverse relation between the pressure and the trace anomaly simplifies to
\be
\frac{p(T,0)}{T^4} = \int_0^T \dd T' \frac{I(T',0)}{T'^5}.
\label{eq:invpi}
\ee
From the trace anomaly and the pressure the energy density $\epsilon$, the entropy density $s$ and the speed of sound $c_s$ are constructed as
\be
\epsilon=I+3p,\quad\;
s = \frac{\epsilon+p-\mu n}{T},\quad\;
c_s^2= \left.\frac{\dd p}{\dd \epsilon}\right|_{s/n}.
\label{eq:thermo}
\ee

Once the entropy density and the particle number density are calculated, one can locate the isentropic trajectories $\mu(T,S/N)$ on the phase diagram, where $s/n=S/N$ is kept constant. The isentropic equation of state is then obtained by expressing the above defined thermodynamic observables as functions of $T$ and $S/N$, e.g.,
\be
p(T,S/N) = p(T,\mu(T,S/N)).
\ee

\section{Simulation and analysis details}
\label{sec:simdet}

On the lattice the temperature and the volume of the system are given by
\be
V=(a N_s)^3,\quad\quad T=(a N_t)^{-1},
\ee
where $a$ is the lattice spacing and $N_s$ ($N_t$) is the number of lattice sites in the spatial (temporal) direction. The continuum limit $a\to0$ at a given temperature is defined by extrapolating to the limit of large temporal extensions $N_t$. 

In this work we use the tree-level improved Symanzik gauge action and a stout smeared staggered fermionic action. The masses of the quarks are set to their physical values, along the line of constant physics (LCP): $m_{ud}(g^2)$ and $m_s(g^2)$, which is determined by keeping the ratios $M_K/f_K$ and $M_\pi/f_K$ at their physical values. 
This LCP corresponds to a constant ratio $m_{s}/m_{ud}=28.15$~\cite{Borsanyi:2010cj}. The lattice scale $a(g^2)$ is fixed using $f_K$. The details of the lattice action and the determination of the LCP and the scale can be found in Refs.~\cite{Aoki:2005vt,Borsanyi:2010cj}.

We use the thermodynamic observables at zero chemical potential, and the leading Taylor-coefficients of the pressure to determine the equation of state at nonzero $\mu$. For the $\mu=0$ equation of state, in Ref.~\cite{Borsanyi:2010cj} we simulated using lattices with $N_t=6,8,10$ and for some temperatures even with $N_t=12$. Since the $N_t=8$ and $10$ results were found to fall on top of each other we gave a continuum estimate based on these two lattice spacings. We use the term `continuum estimate' to express that here an $a\to0$ extrapolation (using at least three lattice spacings and showing that they are in the scaling regime) was not performed.\footnote{See Ref.~\cite{Borsanyi:2012vn} for preliminary results of the continuum extrapolated trace anomaly at $\mu=0$. This extrapolation barely differs from the estimate given in Ref.~\cite{Borsanyi:2010cj}}
 For the study of the fluctuations $\chi_2$ the lattice ensembles were extended in Ref.~\cite{Borsanyi:2011sw} to include $N_t=12$ and $16$ configurations and these were used to extrapolate to the continuum limit. 
Here we employ the entire ensemble of Ref.~\cite{Borsanyi:2011sw} to determine how the EoS is modified due to a nonzero chemical potential. The details of the ensemble parameters can be find in Ref.~\cite{Borsanyi:2011sw}.

The central quantity in our approach is the Taylor-coefficient $\chi_2$ of the pressure.
We consider the continuum extrapolated results of Ref.~\cite{Borsanyi:2011sw} and perform a spline fit to interpolate the data in the temperature range $125 \textmd{ MeV} <T < 400 \textmd{ MeV}$. The systematic error of the so obtained interpolation is determined by varying the position of the nodepoints of the spline function and is added to the statistical errors in quadrature. 
For the $\mu=0$ results reported in Ref.~\cite{Borsanyi:2010cj} we carry out a similar interpolation.
Finally we combine the interpolated continuum $\chi_2$ curve with the interpolated $\mu=0$ observables to obtain the $\mu>0$ pressure and trace anomaly, according to Eqs.~(\ref{eq:pmu}) and~(\ref{eq:imu}). Consequently, at $\mu=0$ the results for all observables presented here are identical to those reported in Ref.~\cite{Borsanyi:2010cj}.
At low temperatures $T\lesssim 125 \textmd{ MeV}$ we use the Hadron Resonance Gas (HRG) model to calculate the observables at both $\mu=0$ and $\mu>0$. The latter -- just as the lattice results -- are obtained via a leading order expansion in $\mu^2$. The detailed description of our implementation of the HRG model can be found in, e.g., Ref.~\cite{Borsanyi:2010cj}.

To estimate the validity region of our Taylor-expansion we measure the next-to-leading order coefficient $\chi_4$ at one lattice spacing ($N_t=8$) and compare its magnitude to the leading order term. We observe that $\chi_4$ gives substantial contribution around the transition temperature, while its weight decreases as $T$ grows. Specifically, we find for the pressure that
\be
\frac{p \textmd{ up to }\mathcal{O}((\mu_B/T)^4)}{p \textmd{ up to }\mathcal{O}((\mu_B/T)^2)} \le 
\begin{cases}
1.1, \hspace*{0.31cm} \textmd{ for $\mu_B/T\le2$}, \\
1.35, \hspace*{0.14cm} \textmd{ for $\mu_B/T\le3$}.
\end{cases}
\label{eq:c4contr}
\ee

At infinitely high temperatures thermodynamic observables approach their corresponding Stefan-Boltz-mann limits. This limit for the pressure for three flavors of quarks is given by
\be
p^{\rm SB}/T^4 = p_0,\quad\quad p_0=19\pi^2/36,
\ee
independently of the chemical potential. At finite entropy over particle number the Stefan-Boltzmann limit is increased, to leading order with $(S/N)^{-2}$, see appendix~\ref{app:SB}.

\section{Results at constant \boldmath \texorpdfstring{$\mu$}{mu}}

\begin{figure}[ht!]
\centering
\includegraphics*[width=8.5cm]{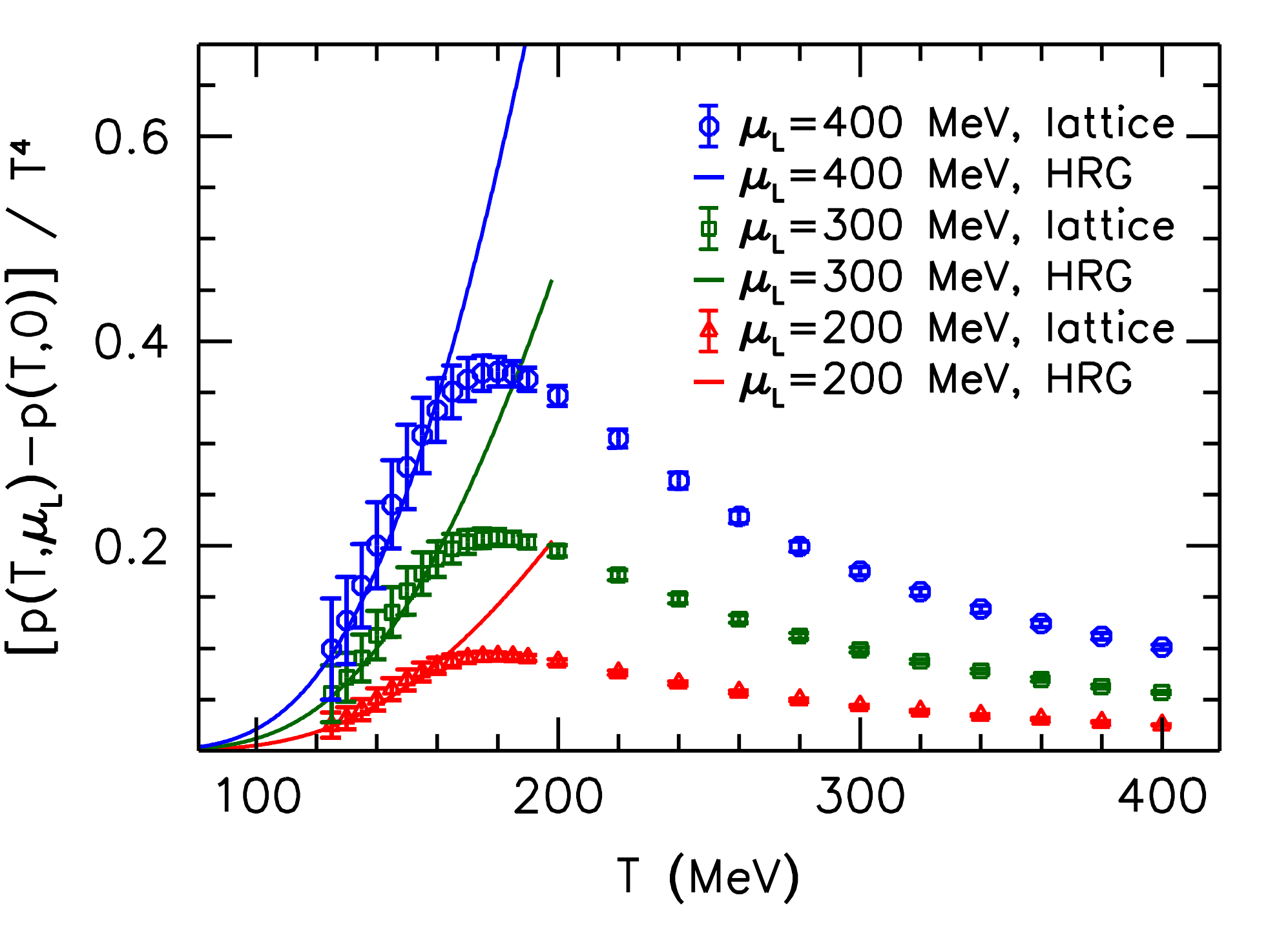}\hspace*{.5cm}
\includegraphics*[width=8.5cm]{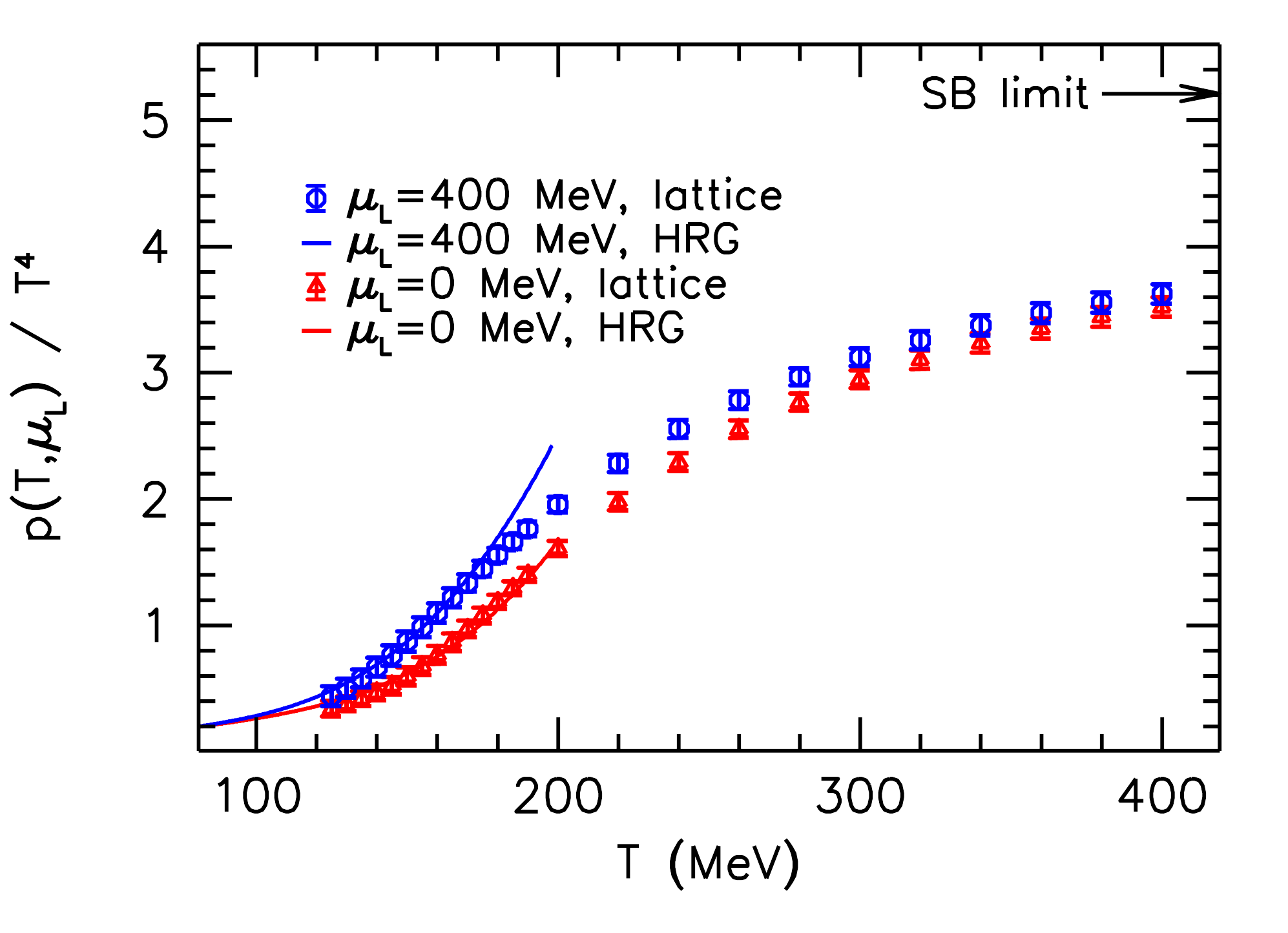}
\vspace*{-0.15cm}
\caption{The difference between the pressure at $\mu>0$ and at $\mu=0$ (left panel). The pressure for nonzero $\mu$, as a function of $T$ (right panel).}
\label{fig:deltap}
\end{figure}

\begin{figure}[ht!]
\centering
\includegraphics*[width=8.5cm]{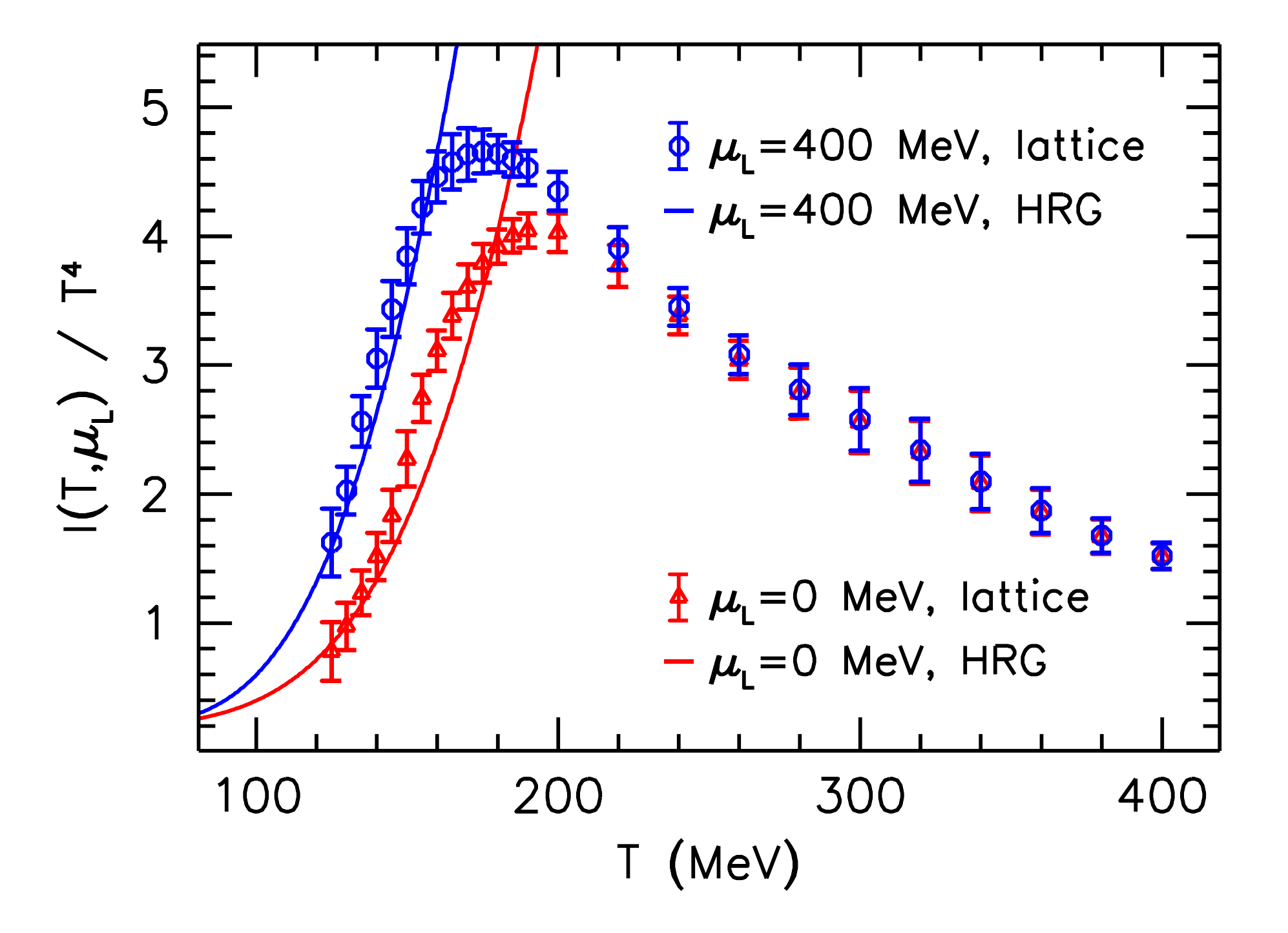}\hspace*{.5cm}
\includegraphics*[width=8.5cm]{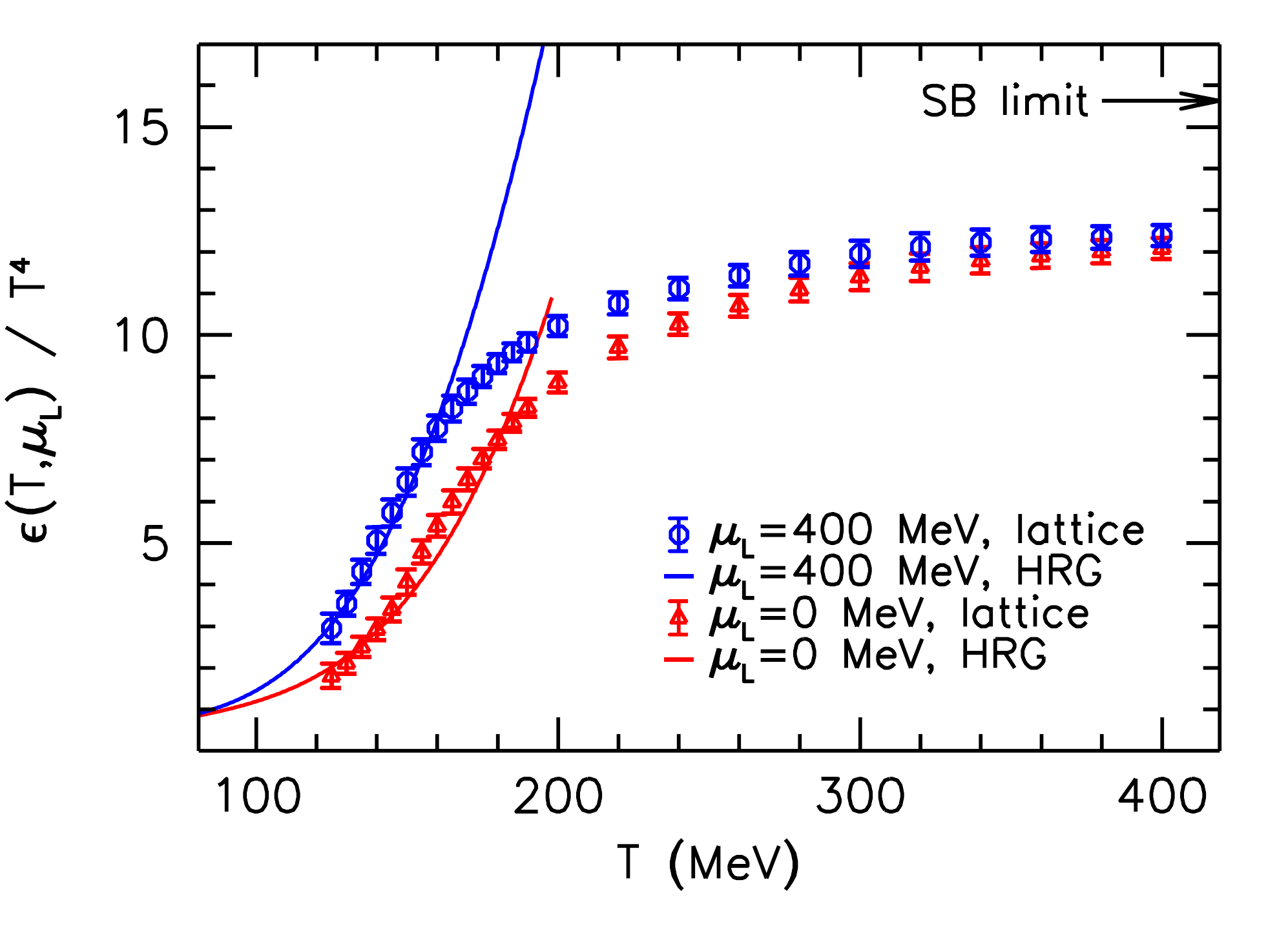}
\vspace*{-0.15cm}
\caption{The trace anomaly (let panel) and the energy density (right panel) for nonzero $\mu$, as functions of $T$.}
\label{fig:imu}
\end{figure}

\begin{figure}[ht!]
\centering
\includegraphics*[width=8.5cm]{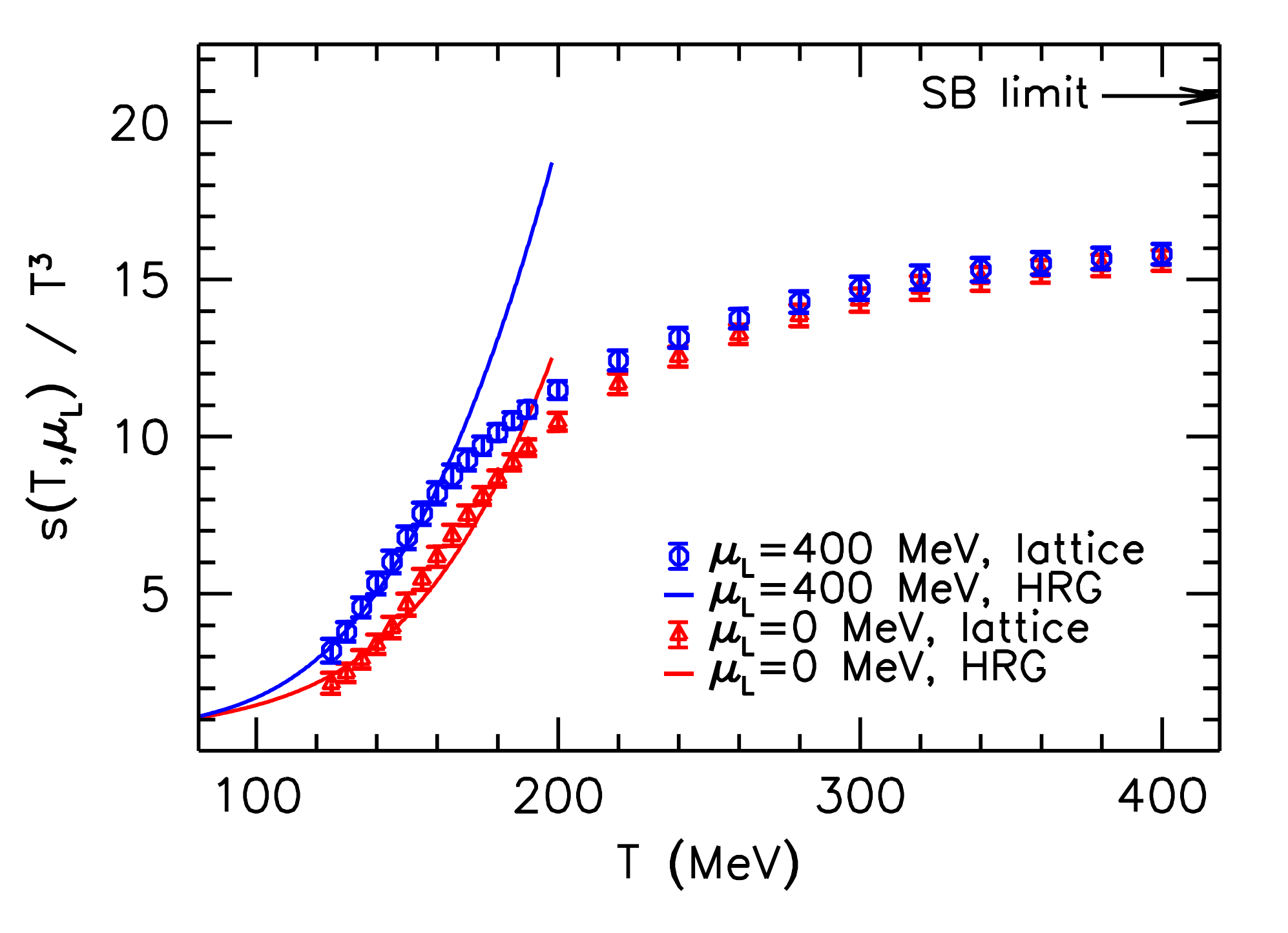}\hspace*{.5cm}
\includegraphics*[width=8.5cm]{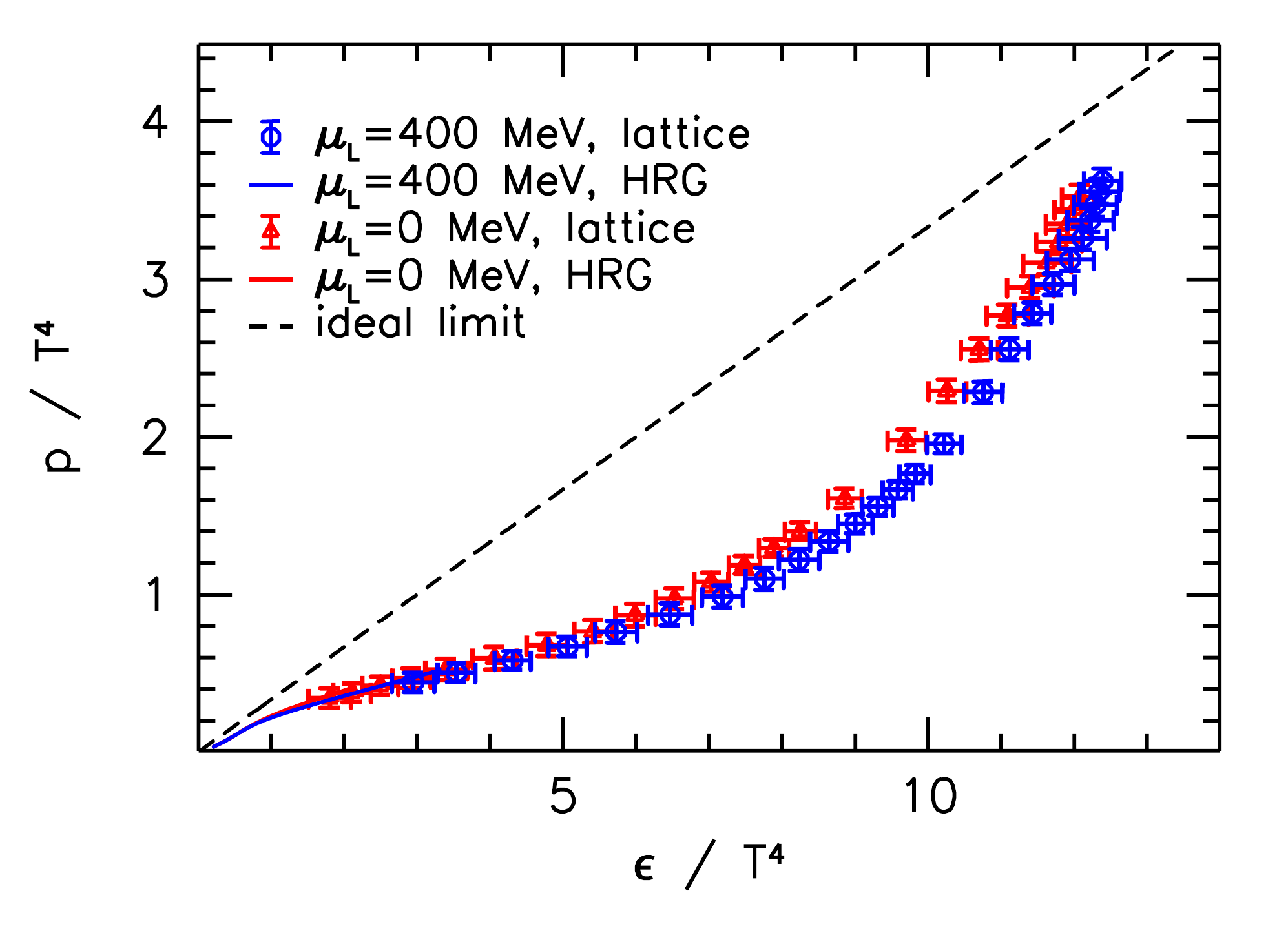}
\vspace*{-0.15cm}
\caption{The entropy density for nonzero $\mu$, as a function of $T$ (left panel). The equation of state for nonzero $\mu$ (right panel).}
\label{fig:smu}
\end{figure}

Using the formulae of section~\ref{sec:formulation}, in particular Eqs.~(\ref{eq:pmu}),~(\ref{eq:imu}) and~(\ref{eq:thermo}) we measure thermodynamic observables as functions of $T$ and $\mu$. For the description of heavy ion collisions the use of the `light' chemical potential $\mu_L$ is preferred (see Eq.~(\ref{eq:muL}) and the remark after). In Figs.~\ref{fig:deltap}-\ref{fig:smu} we plot our continuum estimates for the pressure, the trace anomaly, the energy density and the entropy density. The Stefan-Boltzmann limits of the observables (where different from zero) are shown by the arrows in the upper right corner of the corresponding figures.

\begin{figure}[ht!]
\includegraphics*[width=8.5cm]{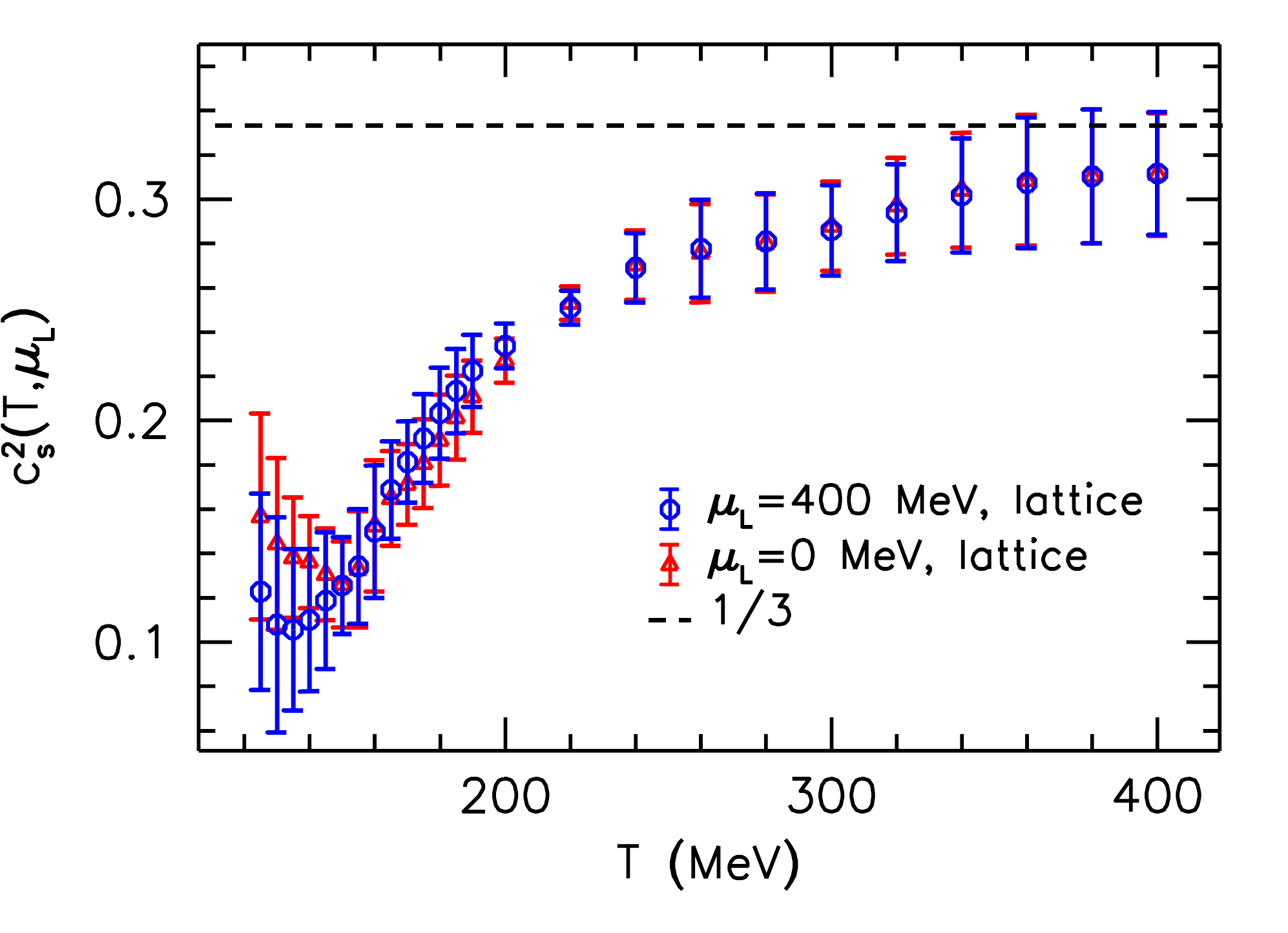}\hspace*{.5cm}
\includegraphics*[width=8.5cm]{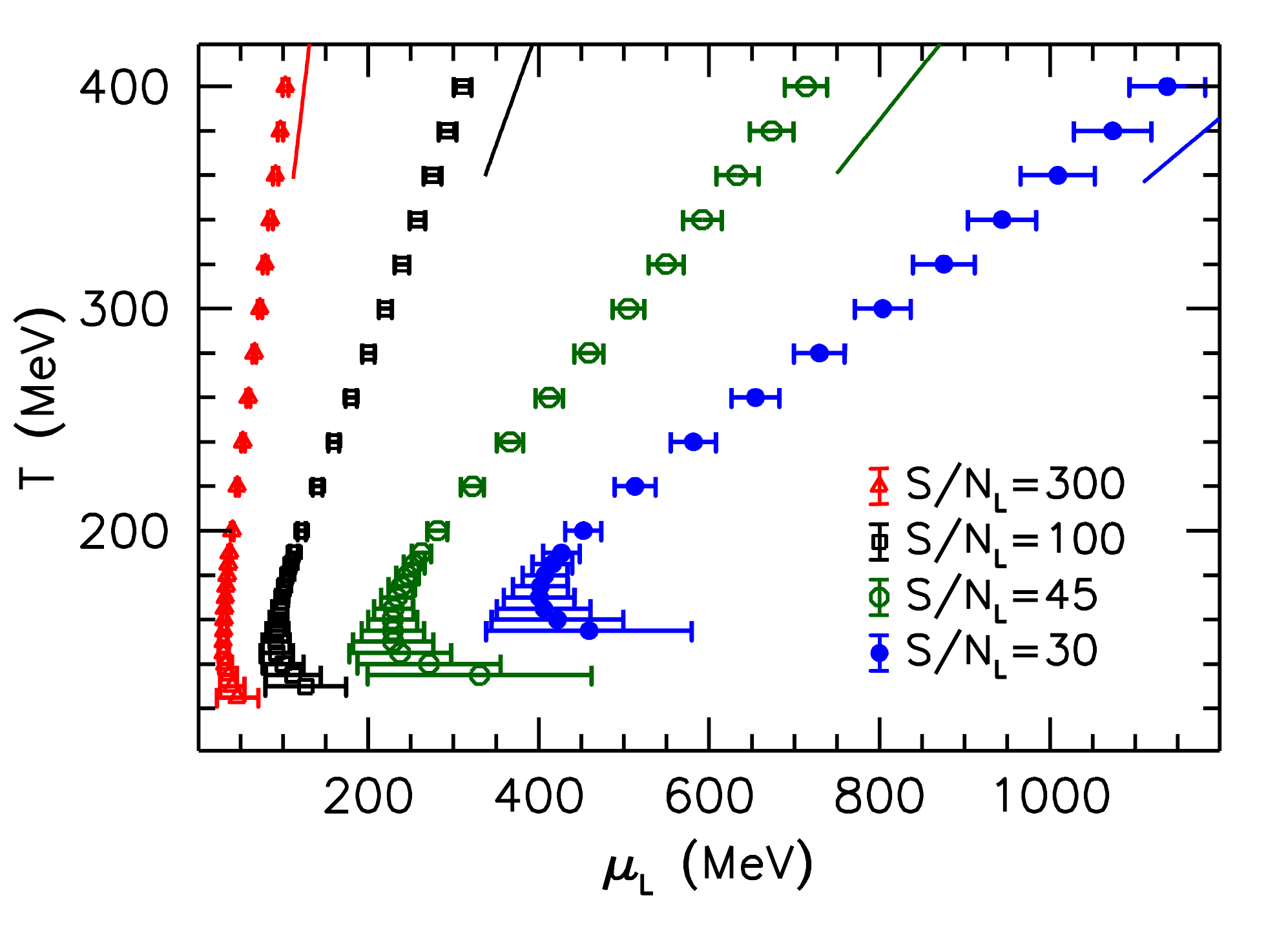}
\vspace*{-0.15cm}
\caption{The square of the speed of sound for nonzero $\mu$, as a function of $T$ (left panel). Trajectories of constant $S/N_L$ on the phase diagram. The lines in the upper part of the plot represent the leading order perturbative expressions (right panel).}
\label{fig:c_s}
\end{figure}

The results for $p/T^4$, $I/T^4$ and $c_s^2$ are also tabulated separately in table~\ref{tab:results} of appendix~\ref{app:tables}. Furthermore, table~\ref{tab:results2} contains the same results for the case of the full baryonic chemical potential, given in Eq.~(\ref{eq:muB}). As expected, the chemical potential increases these observables in the transition region, thus pushing $T_c(\mu)$ towards smaller temperatures for growing $\mu$. For low temperatures we also show the corresponding prediction of the HRG model. 
For all observables, we find a good agreement between these predictions and the lattice data, for temperatures below the transition.

In the right panel of Fig.~\ref{fig:smu} we plot the pressure as a function of the energy density to visualize the QCD equation of state for nonzero chemical potentials $\mu_L>0$. Finally in the left panel of Fig.~\ref{fig:c_s} we plot the square of the speed of sound $c_s^2$ as a function of $T$.  

\section{Results at constant \boldmath \texorpdfstring{$S/N$}{SN}}

Next we locate trajectories of constant $S/N$ on the phase diagram, and present thermodynamic observables along these isentropic lines. We consider $S/N_L=300$, $45$ and $30$, which correspond to typical ratios at RHIC, SPS and AGS~\cite{Bluhm:2007nu} and have been studied in the $N_f=2$~\cite{Ejiri:2005uv} and in the $N_f=2+1$ theory~\cite{Bernard:2007nm,DeTar:2010xm}. These trajectories (together with an intermediate $S/N_L=100$) are plotted in the right panel of Fig.~\ref{fig:c_s}.

\begin{figure}[ht!]
\centering
\includegraphics*[width=8.5cm]{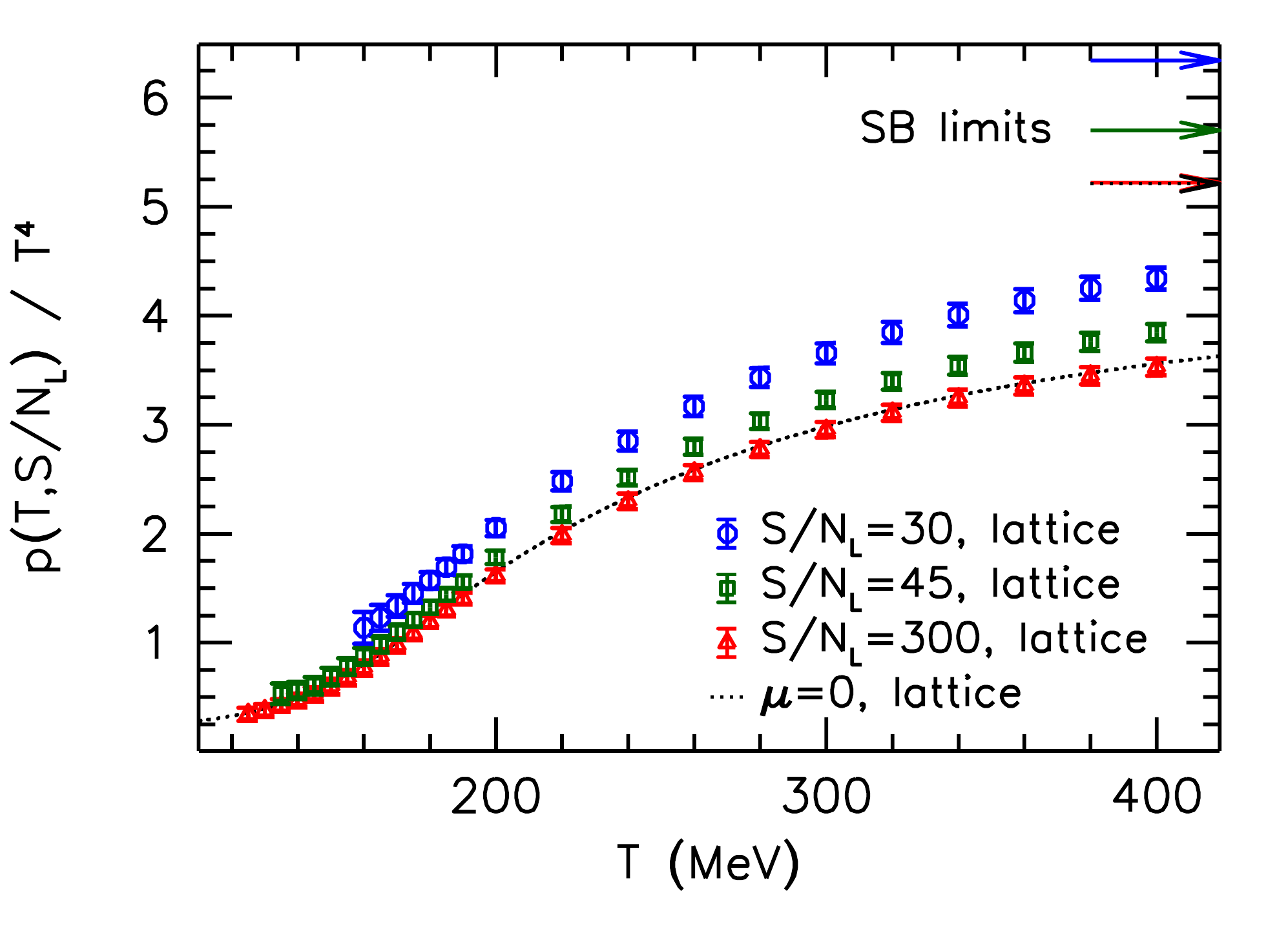}\hspace*{.5cm}
\includegraphics*[width=8.5cm]{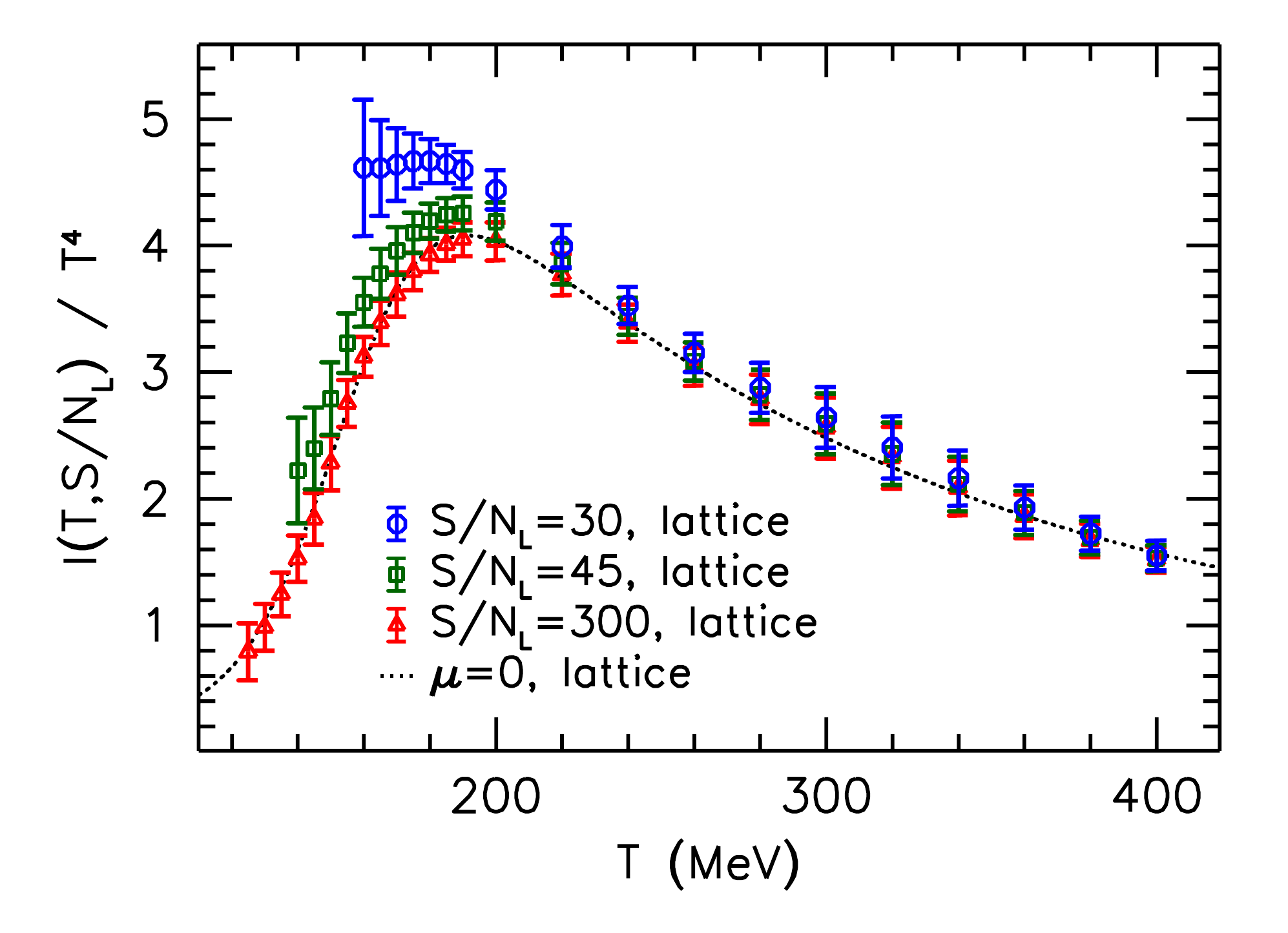}
\vspace*{-0.15cm}
\caption{The pressure (left panel) and the trace anomaly (right panel) as functions of $T$, for various values of the ratio $S/N_L$.}
\label{fig:pmuSNB}
\end{figure}

\begin{figure}[ht!]
\centering
\includegraphics*[width=8.5cm]{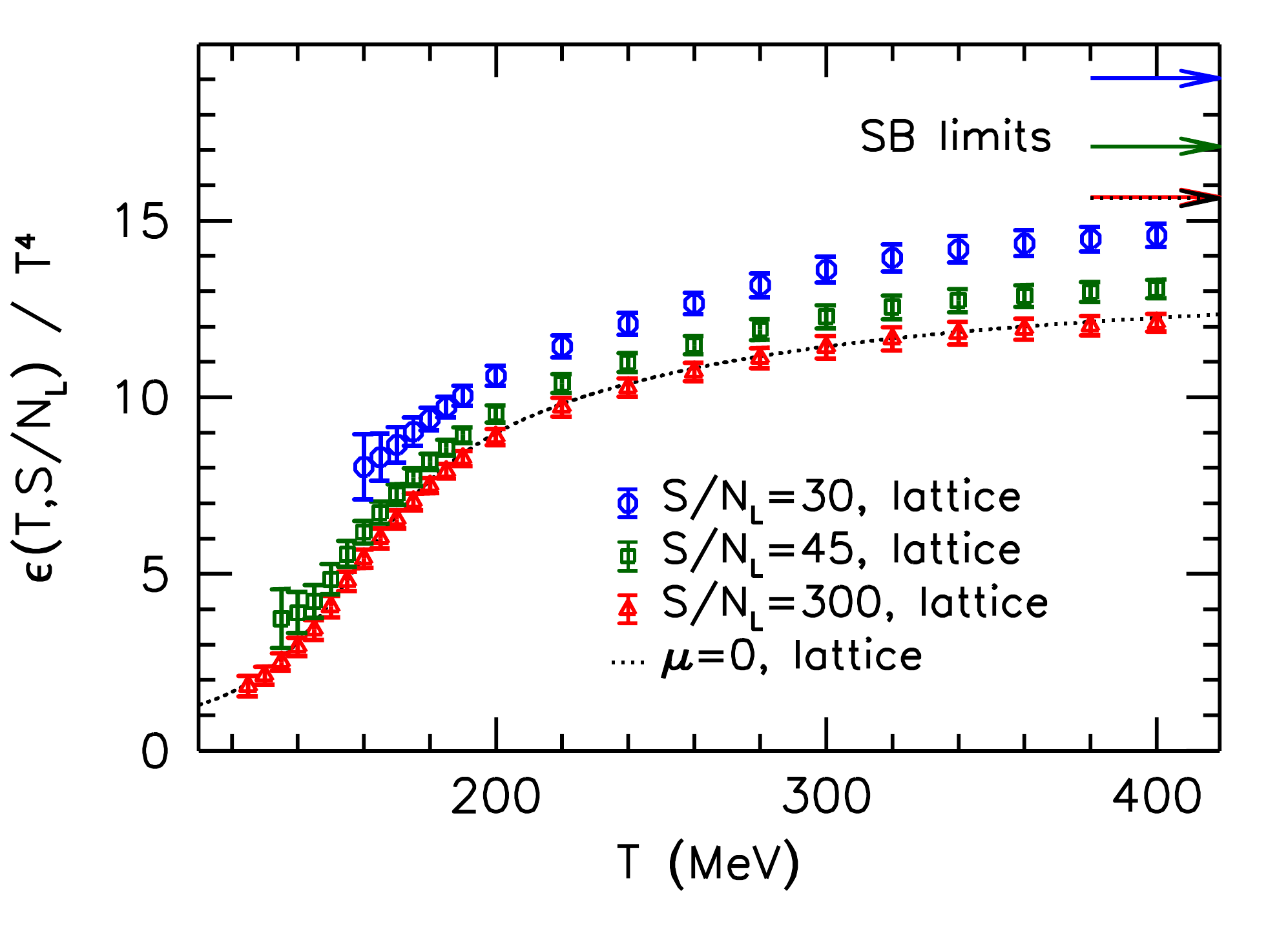}\hspace*{.5cm}
\includegraphics*[width=8.5cm]{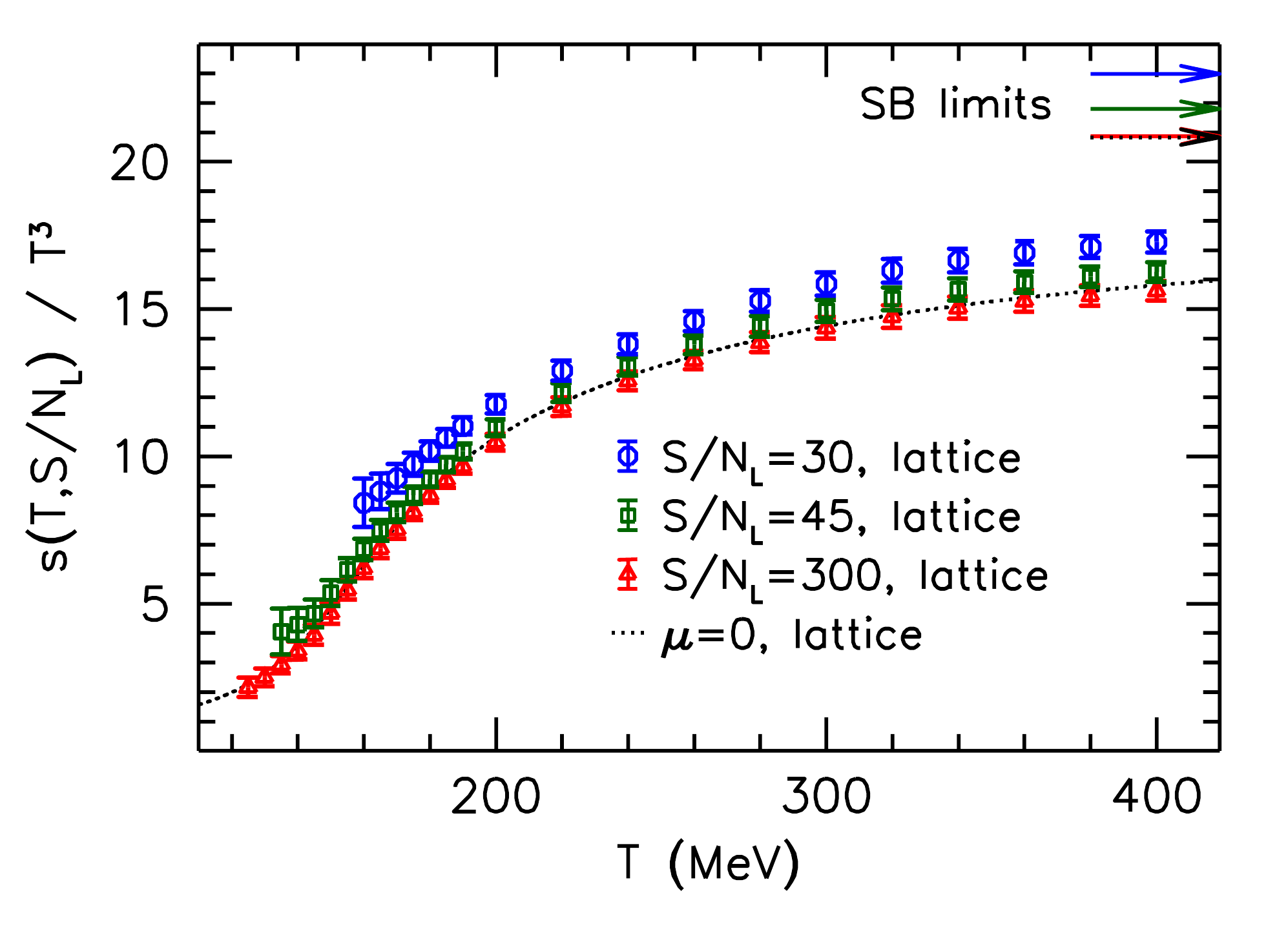}
\vspace*{-0.15cm}
\caption{The energy density (left panel) and the entropy density (right panel) as functions of $T$, for various values of the ratio $S/N_L$.}
\label{fig:emuSNB}
\end{figure}

\begin{figure}[ht!]
\centering
\includegraphics*[width=8.5cm]{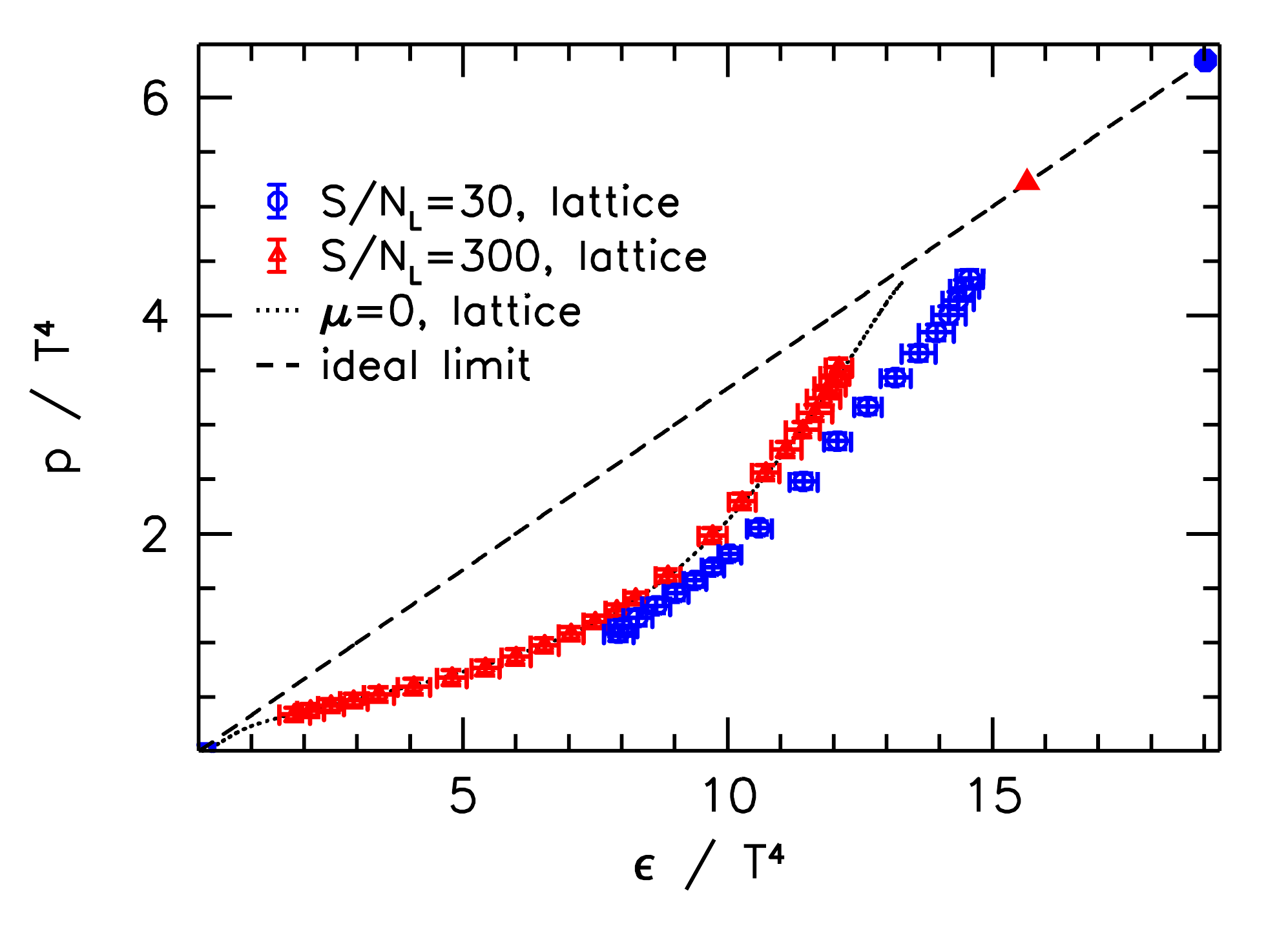}\hspace*{.5cm}
\includegraphics*[width=8.5cm]{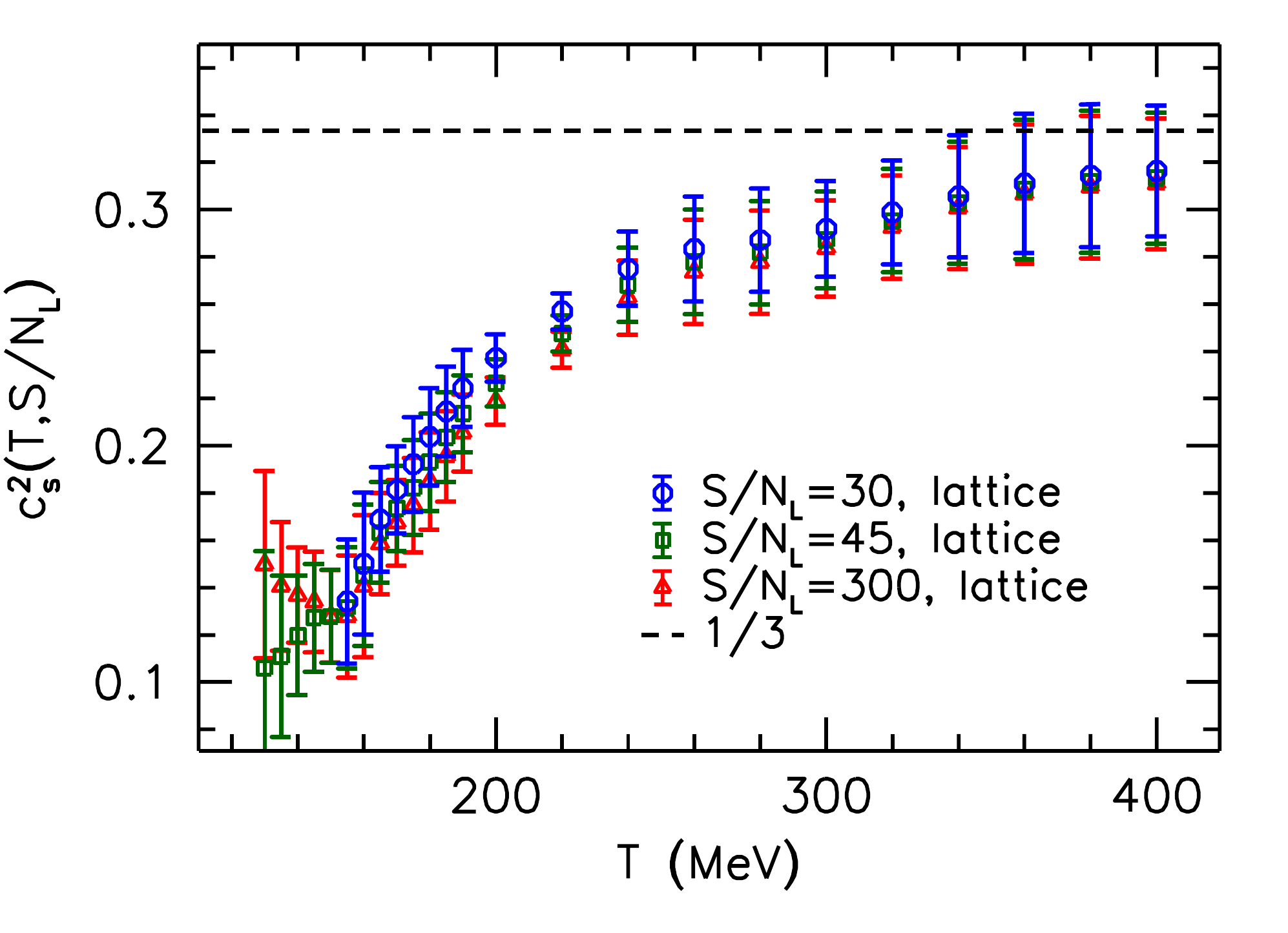}
\vspace*{-0.15cm}
\caption{The equation of state for two values of the ratio $S/N_L$. The dots in the upper right corner represent the Stefan-Boltzmann limits (left panel). The square of the speed of sound as a function of $T$, for various values of the ratio $S/N_L$ (right panel).}
\label{fig:eosSNB}
\end{figure}

At low temperatures $T\lesssim 130 \textmd{ MeV}$ and low values of $S/N_L$ we find that the chemical potential necessary to keep $S/N_L$ constant increases beyond the applicability region $\mu_L(T,S/N_L) < 3T$ of the Taylor-expansion method. Therefore in this case we do not compare the lattice results to the HRG prediction. In Figs.~\ref{fig:pmuSNB}-\ref{fig:eosSNB} we show thermodynamic observables at constant $S/N_L$. We also indicate in the figures the Stefan-Boltzmann limits of the observables for each value of $S/N_L$, see appendix~\ref{app:SB}.

We see that the $S/N_L=300$ results agree with the $\mu=0$ data within errors for all observables, similarly as was observed in Refs.~\cite{Ejiri:2005uv,Bernard:2007nm,DeTar:2010xm}. At $S/N_L=45$ and $30$, on the other hand, the difference to the $\mu=0$ case is quite pronounced for almost all observables.

\section{Parameterization of the results}

We consider the following global parameterization of the results. At $\mu=0$ the trace anomaly was fitted in Ref.~\cite{Borsanyi:2010cj} by
\be
\frac{I(T)}{T^4}\!=
e^{-h_1/t - h_2/t^2}\!\cdot \!\left[ h_0 + \frac{f_0\!\cdot\! [\tanh(f_1\cdot t+f_2)+1]}{1+g_1\cdot t+g_2\cdot t^2} \right]\!,
\label{eq:pari}
\ee
while the Taylor-coefficient we choose to parameterize by
\be
\chi_2(T)= e^{-h_3/t - h_4/t^2}\cdot f_3\cdot [\tanh(f_4\cdot t+f_5)+1],
\label{eq:parc2}
\ee
\begin{wrapfigure}{r}{8.9cm}
\vspace*{-0.2cm}
\includegraphics*[width=8.5cm]{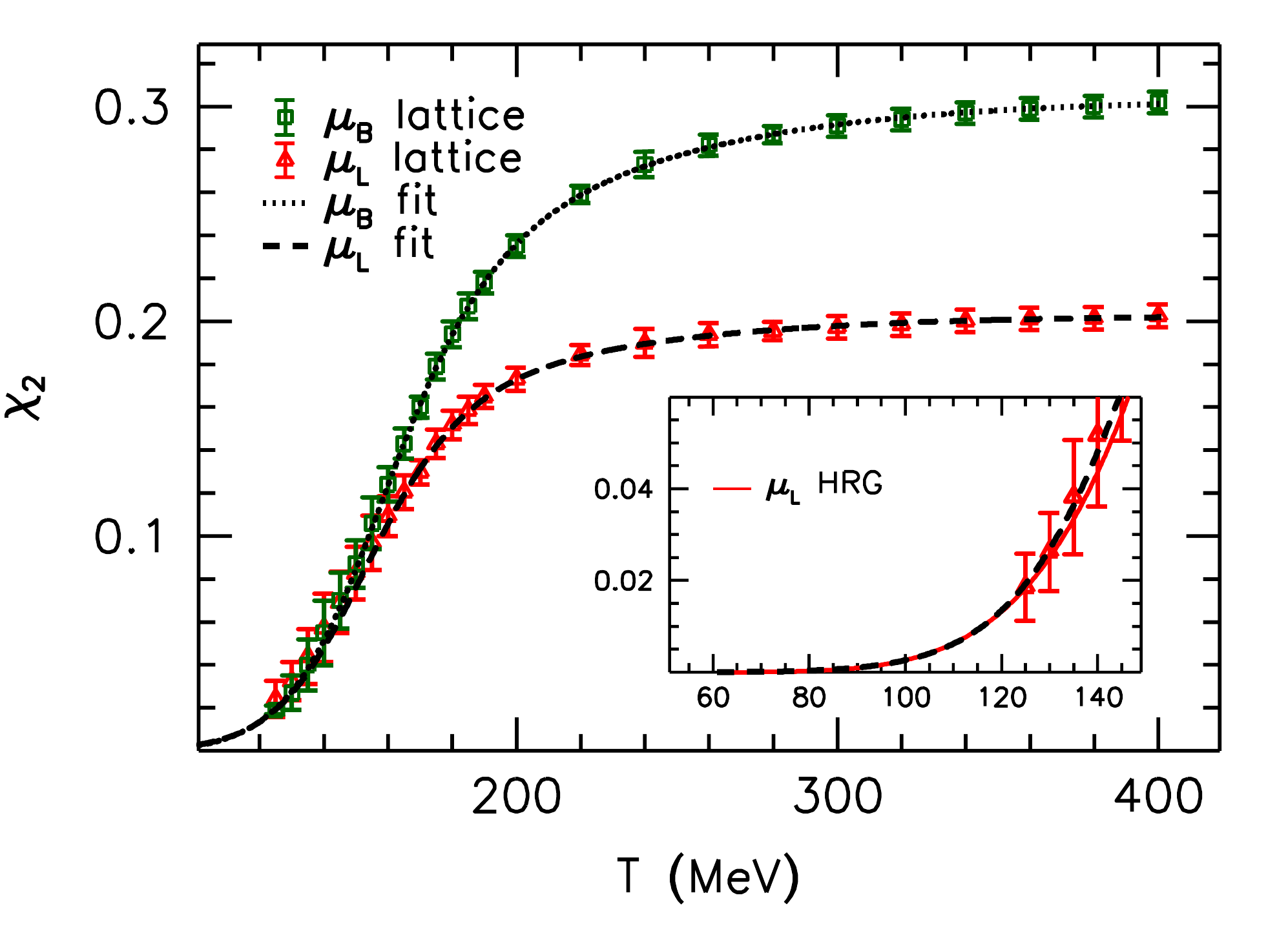}
\vspace*{-0.15cm}
\caption{Parameterization of the lattice results and the HRG prediction for $\chi_2$, for $\mu_L$ and $\mu_B$. The inset zooms into the low temperature region (here results for $\mu_B$ and for $\mu_L$ are on top of each other, so we only plot the latter).}
\label{fig:param}
\end{wrapfigure}
where $t=T/200 \textmd{ MeV}$. We show the fit parameters for $I/T^4$ and $\chi_2$ in table~\ref{tab:pars}. For the latter case we tabulate the parameters for both choice of the chemical potential (Eqs.~(\ref{eq:muL}) and~(\ref{eq:muB})). These functions reproduce for $T>125 \textmd{ MeV}$ the lattice data for the trace anomaly and for the Taylor-coefficient, respectively. Below this temperature they smoothly interpolate to the HRG data, and can be used to calculate the observables in the whole temperature window $0<T<400 \textmd{ MeV}$. The fits for the Taylor-coefficients for $\mu_L$ and $\mu_B$ are shown in Fig.~\ref{fig:param}.
The deviation between the fit for $\chi_2$ and the lattice results is under $0.004$ in both cases. Moreover, the difference between the fit and the HRG prediction is smaller than $0.001$ up to $T\le 120 \textmd{ MeV}$.

The pressure at zero chemical potential can be obtained from the $\mu=0$ trace anomaly through the definite integral~(\ref{eq:invpi}). Combined with the parameterization for $\chi_2$ the pressure at nonzero $\mu$ is obtained by Eq.~(\ref{eq:pmu}). The trace anomaly at nonzero $\mu$ can be calculated simply using Eq.~(\ref{eq:imu}) where the derivative of $\chi_2$ with respect to the temperature enters. The energy density and the entropy density at nonzero $\mu$ are obtained via Eq.~(\ref{eq:thermo}). To obtain the isentropic EoS, one can numerically locate the trajectories of constant $S/N$ and express the observables as functions of $T$.
Considering the speed of sound; as visible in Figs.~\ref{fig:c_s} and~\ref{fig:eosSNB}, the $\mu$-dependence (or the $S/N$-dependence) of $c_s^2$ cannot be established from the lattice data at the present level of statistics. 
Therefore a good description of $c_s^2$ can be obtained in terms of the $\mu=0$ parameterization, as derived from Eq.~(\ref{eq:pari}).
The parameterization of the observables (both as functions of $T,\mu$ and as functions of $T,S/N$) is implemented in a Matlab/Octave script ({\tt parameterization.m}) that is submitted to the arXiv as ancillary file along with this paper.

\begin{table}[ht!]
\centering
\begin{tabular}{|c|c|c|c|c|c|c|c|}
\hline
$h_0$ & $h_1$ & $h_2$ & $f_0$ & $f_1$ & $f_2$ & $g_1$ & $g_2$ \\
\hline
0.1396 & -0.1800 & 0.0350 & 2.76 & 6.79 & -5.29 & -0.47 & 1.04\\
\hline
\end{tabular}\\[0.1cm]
\begin{tabular}{|c|c|c|c|c|c|}
\hline
$\mu$ & $h_3$ & $h_4$ & $f_3$ & $f_4$ & $f_5$ \\
\hline
$\mu_L$ & 
-0.3364 & 0.3902 & 0.0940 & 6.8312 & -5.0907 \\ \hline
$\mu_B$ &
-0.5022 & 0.5950 & 0.1359 & 6.3290 & -4.8303 \\
\hline
\end{tabular}

\caption{\label{tab:pars}
Parameters of the functions~(\ref{eq:pari}) and~(\ref{eq:parc2}).}
\end{table}

Using this parameterization we determine the characteristic points $T_{\rm ch}$ of the above presented observables. 
The dependence of $T_{\rm ch}$ on $\mu^2$ defines the curvature $\kappa$ of the transition line at $\mu=0$,
\be
T_{\rm ch}(\mu) = T_{\rm ch}(0) \left[ 1 - \kappa \frac{\mu^2}{T_{\rm ch}(0)^2} \right].
\ee
In table~\ref{tab:tcs} we show the results for the curvature from different definitions. We define $T_{\rm ch}$ using the inflection point/maximum of either the trace anomaly, the energy density, the entropy density or the ratio $p/\epsilon$.

\begin{table}[ht!]
\centering
\begin{tabular}{|c|c||c|c|}
\hline
definition & $\kappa$ & definition & $\kappa$ \\
\hline \hline
$I/T^4$ inf. & 0.015(6) & $I/T^4$ max. & 0.020(2) \\
\hline
$\epsilon/T^4$ inf. & 0.017(5) & $s/T^3$ inf. & 0.016(4) \\
\hline
$p/\epsilon$ inf. & 0.018(7) & & \\
\hline
\end{tabular}
\caption{\label{tab:tcs}
Curvatures from different definitions, in terms of inflection points (inf.) or maxima (max.) of the corresponding observables.}
\end{table}

These results are somewhat larger than our previous findings determined from inflection points of the strange quark number susceptibility, $\kappa=0.0089(14)$, or of the chiral condensate $\kappa=0.0066(20)$ in the continuum limit~\cite{Endrodi:2011gv}.
Similar results have beein obtained for the curvature using either Taylor expansion, imaginary chemical potentials or the canonical ensemble, for a review see, e.g. Ref.~\cite{Fodor:2009ax}.
We remark that differences between curvatures using different definitions can be expected due to the crossover nature of the transition.

\section{Summary}

In this paper we determined the QCD equation of state for small chemical potentials using a Taylor-expansion technique. We employed $2+1$ flavors of quarks with physical masses and estimated the continuum limit of our data using lattices with $N_t=6$, 8, 10, 12 and 16. We presented results regarding various thermodynamic observables as functions of the temperature and the `light' baryonic chemical potential $\mu_L$, which is the relavant parameter for the description of heavy ion collisions with zero net strangeness density. We also determined the isentropic equation of state and showed the observables along lines of constant entropy over particle number.
A global parameterization of our observables is given such that they can be reconstructed for small chemical potentials $\mu_L/T_c=3 \mu_u/T_c\lesssim 3$ in the temperature window $0<T<400 \textmd{ MeV}$.
Note that since our observables are calculated via a leading order ($\mathcal{O}(\mu^2)$) Taylor-expansion, one must consider truncation errors for moderate chemical potentials. We find that these are largest in the transition region, see Eq.~(\ref{eq:c4contr}).
We also used inflection points/maxima of our observables to define the characteristic temperatures as functions of the chemical potential and to determine the corresponding curvature $\kappa$ for each definition. Our results are relevant for contemporary and upcoming heavy ion collision experiments where the low $\mu$ - high $T$ region of the QCD phase diagram is explored.

\begin{acknowledgments}
This research has been partly supported by the Research Executive Agency (REA) of the European Union under Grant Agreement number PITN-GA-2009-238353 (ITN STRONGnet) and the European Research Council grant 208740 (FP7/2007-2013), as well as the Italian Ministry of Education, Universities and Research under the Firb Research Grant RBFR0814TT.
Computations were performed on the Blue Gene supercomputers at FZ J\"ulich, on the QPACE facility supported by the Sonderforschungsbereich TR55 and on GPU~\cite{Egri:2006zm} clusters at Wuppertal and also at the E\"otv\"os University, Budapest.
\end{acknowledgments}

\bibliographystyle{JHEP}
\bibliography{eosmu}

\appendix
\section{Stefan-Boltzmann limits}\label{app:SB}

To lowest order in perturbation theory, the pressure for three flavors of quarks is given by (see, e.g. Ref.~\cite{kapusta1989finite}),
\be
\frac{p^{\rm SB}}{T^4} = \frac{19\pi^2}{36} + \sum_i \left( \frac{1}{2}\frac{\mu_i^2}{T^2} + \frac{1}{4\pi^2} \frac{\mu_i^4}{T^4}   \right).
\label{eq:pSB}
\ee
Let us consider the light baryonic chemical potential, Eq.~(\ref{eq:muL}). In the high temperature limit $\mu_s$ approaches zero since $\chi_2^{us}/\chi_2^{ss}\to0$. Therefore, the expression~(\ref{eq:pSB}) reduces to
\be
\frac{p^{\rm SB}(\mu_L)}{T^4} = p_0 + \frac{1}{9}\frac{\mu_L^2}{T^2} + \frac{1}{162 \pi^2} \frac{\mu_L^4}{T^4}.
\label{eq:pres_SB}
\ee
Consequently, the baryon number density is,
\be
\frac{n^{\rm SB}_L}{T^3} = \frac{2}{9}\frac{\mu_L}{T} + \frac{2}{81 \pi^2} \frac{\mu_L^3}{T^3},
\ee
and the entropy density
\be
\frac{s^{\rm SB}(\mu_L)}{T^3} = 4p_0 + \frac{2}{9}\frac{\mu_L^2}{T^2}.
\label{eq:sSB}
\ee
Solving the equation
\be
s^{SB}/ n^{SB}_L = S/N_L = \textmd{const.}
\ee
for $\mu_L/T$ one obtains the limiting behavior for the isentropic lines on the phase diagram, as depicted by the solid lines in the right panel of Fig.~\ref{fig:c_s}. Putting back the solution in Eqs.~(\ref{eq:pres_SB}) and~(\ref{eq:sSB}) one obtains the Stefan-Boltzmann limit of the pressure and of the entropy density, respectively, for the $S/N_L$ value in question. To leading order in $(S/N_L)^{-2}$ this amounts to
\be
\frac{p^{\rm SB}(S/N_L)}{T^4} = p_0 \cdot\left[1 + \frac{ x_L \cdot p_0}{(S/N_{L})^2}\right], \quad\quad x_L=36.
\ee
For the full baryonic chemical potential, Eq.~(\ref{eq:muB}), a similar calculation gives $x_B=24$.

\section{Tables}
\label{app:tables}
\vspace*{0.5cm}

\setlength{\tabcolsep}{1.5pt}
\setlength{\extrarowheight}{2pt}

\begin{center}
\fontsize{9.5}{9}\selectfont
\mbox{
\input{table_L}
}
\begin{table}[ht!]
\caption{Lattice results for the pressure, the trace anomaly and the speed of sound squared, as functions of the temperature and the `light' baryonic chemical potential.
\label{tab:results}
}

\end{table}
\end{center}

\begin{center}
\fontsize{9.5}{9}\selectfont
\mbox{
\input{table_B}
}
\begin{table}[ht!]
\caption{Lattice results for the pressure, the trace anomaly and the speed of sound squared, as functions of the temperature and the full baryonic chemical potential.
\label{tab:results2}
}
\end{table}
\end{center}

\end{document}

%% file: table_L.tex
\begin{tabular}{|c||c|c|c||c|c|c||c|c|c||c|c|c||c|c|c|}
\hline
\multirow{2}{*}{$T (\textmd{MeV})$} & \multicolumn{3}{|c||}{$\mu_L=0$} 
& \multicolumn{3}{|c||}{$\mu_L=100 \textmd{ MeV}$}
& \multicolumn{3}{|c||}{$\mu_L=200 \textmd{ MeV}$}
& \multicolumn{3}{|c||}{$\mu_L=300 \textmd{ MeV}$}    
& \multicolumn{3}{|c|}{$\mu_L=400 \textmd{ MeV}$}\\
\cline{2-16}
& $p/T^4$ & $I/T^4$ & $c_s^2$ 
& $p/T^4$ & $I/T^4$ & $c_s^2$
& $p/T^4$ & $I/T^4$ & $c_s^2$
& $p/T^4$ & $I/T^4$ & $c_s^2$  
& $p/T^4$ & $I/T^4$ & $c_s^2$ \\
\hline\hline
125 & 0.34(6) & 0.8(2) & 0.16(5)  & 0.35(6) & 0.8(2)  & 0.15(5)  & 0.37(6) & 1.0(2)  & 0.15(5)  & 0.40(7) & 1.3(2)  & 0.13(5)  & 0.44(8) & 1.6(3)  & 0.12(5) \\ 
130 & 0.38(6) & 1.0(2) & 0.14(4)  & 0.39(6) & 1.0(2)  & 0.14(4)  & 0.41(6) & 1.2(2)  & 0.13(4)  & 0.45(6) & 1.6(2)  & 0.12(5)  & 0.51(7) & 2.0(2)  & 0.11(5) \\ 
135 & 0.42(6) & 1.2(2) & 0.14(3)  & 0.43(6) & 1.3(2)  & 0.14(3)  & 0.46(6) & 1.6(2)  & 0.12(3)  & 0.51(6) & 2.0(2)  & 0.11(3)  & 0.58(7) & 2.6(2)  & 0.11(4) \\ 
140 & 0.47(6) & 1.5(2) & 0.14(2)  & 0.48(6) & 1.6(2)  & 0.13(2)  & 0.52(6) & 1.9(2)  & 0.13(2)  & 0.58(7) & 2.4(2)  & 0.12(3)  & 0.67(7) & 3.1(2)  & 0.11(3) \\ 
145 & 0.52(7) & 1.8(2) & 0.13(2)  & 0.54(7) & 1.9(2)  & 0.13(2)  & 0.58(7) & 2.2(2)  & 0.13(2)  & 0.66(7) & 2.7(2)  & 0.12(3)  & 0.76(8) & 3.4(2)  & 0.12(3) \\ 
150 & 0.60(7) & 2.3(2) & 0.13(2)  & 0.61(7) & 2.4(2)  & 0.13(2)  & 0.67(7) & 2.7(2)  & 0.13(2)  & 0.75(7) & 3.2(2)  & 0.13(2)  & 0.87(8) & 3.8(2)  & 0.13(2) \\ 
155 & 0.68(7) & 2.7(2) & 0.13(3)  & 0.70(7) & 2.8(2)  & 0.13(3)  & 0.76(7) & 3.1(2)  & 0.13(3)  & 0.85(7) & 3.6(2)  & 0.13(3)  & 0.99(8) & 4.2(2)  & 0.13(3) \\ 
160 & 0.77(7) & 3.1(2) & 0.15(3)  & 0.79(7) & 3.2(2)  & 0.14(3)  & 0.85(7) & 3.5(2)  & 0.14(3)  & 0.95(7) & 3.9(2)  & 0.15(3)  & 1.10(8) & 4.5(2)  & 0.15(3) \\ 
165 & 0.87(7) & 3.4(2) & 0.16(2)  & 0.89(7) & 3.5(2)  & 0.16(2)  & 0.96(7) & 3.7(2)  & 0.16(2)  & 1.07(7) & 4.1(2)  & 0.17(2)  & 1.22(8) & 4.6(2)  & 0.17(2) \\ 
170 & 0.97(7) & 3.6(2) & 0.17(2)  & 1.00(7) & 3.7(2)  & 0.17(2)  & 1.06(7) & 3.9(2)  & 0.17(2)  & 1.18(7) & 4.2(2)  & 0.18(2)  & 1.34(7) & 4.6(2)  & 0.18(2) \\ 
175 & 1.08(6) & 3.8(2) & 0.18(2)  & 1.10(6) & 3.8(2)  & 0.18(2)  & 1.17(6) & 4.0(2)  & 0.18(2)  & 1.29(6) & 4.3(2)  & 0.19(2)  & 1.45(6) & 4.7(2)  & 0.19(2) \\ 
180 & 1.19(6) & 3.9(1) & 0.19(2)  & 1.21(6) & 4.0(1)  & 0.19(2)  & 1.28(6) & 4.1(1)  & 0.19(2)  & 1.40(6) & 4.3(1)  & 0.20(2)  & 1.56(6) & 4.6(1)  & 0.20(2) \\ 
185 & 1.30(5) & 4.0(1) & 0.20(2)  & 1.32(5) & 4.0(1)  & 0.20(2)  & 1.39(5) & 4.2(1)  & 0.20(2)  & 1.50(6) & 4.3(1)  & 0.21(2)  & 1.66(6) & 4.6(1)  & 0.21(2) \\ 
190 & 1.40(6) & 4.0(1) & 0.21(2)  & 1.43(6) & 4.1(1)  & 0.21(2)  & 1.49(6) & 4.2(1)  & 0.21(2)  & 1.61(6) & 4.3(1)  & 0.22(2)  & 1.77(6) & 4.5(1)  & 0.22(2) \\ 
200 & 1.61(6) & 4.0(2) & 0.23(1)  & 1.63(6) & 4.0(2)  & 0.22(1)  & 1.70(6) & 4.1(2)  & 0.22(1)  & 1.81(6) & 4.2(2)  & 0.23(1)  & 1.96(6) & 4.3(2)  & 0.23(1) \\ 
220 & 1.98(7) & 3.8(2) & 0.25(1)  & 2.00(7) & 3.8(2)  & 0.24(1)  & 2.05(7) & 3.8(2)  & 0.24(1)  & 2.15(7) & 3.8(2)  & 0.25(1)  & 2.28(7) & 3.9(2)  & 0.25(1) \\ 
240 & 2.29(7) & 3.4(1) & 0.27(2)  & 2.31(7) & 3.4(1)  & 0.26(2)  & 2.36(7) & 3.4(1)  & 0.26(2)  & 2.44(7) & 3.4(1)  & 0.27(2)  & 2.56(7) & 3.5(1)  & 0.27(2) \\ 
260 & 2.55(7) & 3.0(1) & 0.28(2)  & 2.57(7) & 3.0(1)  & 0.27(2)  & 2.61(7) & 3.1(1)  & 0.27(2)  & 2.68(7) & 3.1(1)  & 0.28(2)  & 2.78(7) & 3.1(1)  & 0.28(2) \\ 
280 & 2.77(7) & 2.8(2) & 0.28(2)  & 2.78(7) & 2.8(2)  & 0.28(2)  & 2.82(7) & 2.8(2)  & 0.28(2)  & 2.88(7) & 2.8(2)  & 0.28(2)  & 2.97(7) & 2.8(2)  & 0.28(2) \\ 
300 & 2.95(7) & 2.6(2) & 0.29(2)  & 2.96(7) & 2.6(2)  & 0.28(2)  & 2.99(7) & 2.6(2)  & 0.28(2)  & 3.05(7) & 2.6(2)  & 0.28(2)  & 3.12(7) & 2.6(2)  & 0.29(2) \\ 
320 & 3.10(7) & 2.3(2) & 0.30(2)  & 3.11(7) & 2.3(2)  & 0.29(2)  & 3.14(7) & 2.3(2)  & 0.29(2)  & 3.19(7) & 2.3(2)  & 0.29(2)  & 3.26(7) & 2.3(2)  & 0.29(2) \\ 
340 & 3.24(8) & 2.1(2) & 0.30(3)  & 3.25(8) & 2.1(2)  & 0.30(3)  & 3.27(8) & 2.1(2)  & 0.30(3)  & 3.31(8) & 2.1(2)  & 0.30(3)  & 3.37(8) & 2.1(2)  & 0.30(3) \\ 
360 & 3.35(8) & 1.9(2) & 0.31(3)  & 3.36(8) & 1.9(2)  & 0.31(3)  & 3.38(8) & 1.9(2)  & 0.31(3)  & 3.42(8) & 1.9(2)  & 0.31(3)  & 3.47(8) & 1.9(2)  & 0.31(3) \\ 
380 & 3.44(8) & 1.7(1) & 0.31(3)  & 3.45(8) & 1.7(1)  & 0.31(3)  & 3.47(8) & 1.7(1)  & 0.31(3)  & 3.51(8) & 1.7(1)  & 0.31(3)  & 3.56(8) & 1.7(1)  & 0.31(3) \\ 
400 & 3.52(8) & 1.5(1) & 0.31(3)  & 3.53(8) & 1.5(1)  & 0.31(3)  & 3.55(8) & 1.5(1)  & 0.31(3)  & 3.58(8) & 1.5(1)  & 0.31(3)  & 3.62(8) & 1.5(1)  & 0.31(3) \\ 
\hline
\end{tabular}

%% file: table_B.tex
\begin{tabular}{|c||c|c|c||c|c|c||c|c|c||c|c|c||c|c|c|}
\hline
\multirow{2}{*}{$T (\textmd{MeV})$} & \multicolumn{3}{|c||}{$\mu_B=0$} 
& \multicolumn{3}{|c||}{$\mu_B=100 \textmd{ MeV}$}
& \multicolumn{3}{|c||}{$\mu_B=200 \textmd{ MeV}$}
& \multicolumn{3}{|c||}{$\mu_B=300 \textmd{ MeV}$}    
& \multicolumn{3}{|c|}{$\mu_B=400 \textmd{ MeV}$}\\
\cline{2-16}
& $p/T^4$ & $I/T^4$ & $c_s^2$ 
& $p/T^4$ & $I/T^4$ & $c_s^2$
& $p/T^4$ & $I/T^4$ & $c_s^2$
& $p/T^4$ & $I/T^4$ & $c_s^2$  
& $p/T^4$ & $I/T^4$ & $c_s^2$ \\
\hline\hline
125 & 0.34(6) & 0.8(2) & 0.16(5)  & 0.35(6) & 0.8(2)  & 0.15(5)  & 0.37(6) & 1.0(2)  & 0.15(5)  & 0.40(6) & 1.3(2)  & 0.14(5)  & 0.44(6) & 1.7(3)  & 0.14(5) \\ 
130 & 0.38(6) & 1.0(2) & 0.14(4)  & 0.39(6) & 1.0(2)  & 0.15(4)  & 0.41(6) & 1.3(2)  & 0.14(4)  & 0.45(6) & 1.6(2)  & 0.13(4)  & 0.51(6) & 2.1(3)  & 0.12(4) \\ 
135 & 0.42(6) & 1.2(2) & 0.14(3)  & 0.43(6) & 1.3(2)  & 0.14(3)  & 0.46(6) & 1.6(2)  & 0.13(3)  & 0.52(6) & 2.0(2)  & 0.12(4)  & 0.59(6) & 2.6(3)  & 0.11(4) \\ 
140 & 0.47(6) & 1.5(2) & 0.14(2)  & 0.48(6) & 1.6(2)  & 0.13(2)  & 0.52(6) & 1.9(2)  & 0.13(2)  & 0.59(6) & 2.5(2)  & 0.12(3)  & 0.68(7) & 3.2(3)  & 0.11(4) \\ 
145 & 0.52(7) & 1.8(2) & 0.13(2)  & 0.54(7) & 2.0(2)  & 0.13(2)  & 0.59(7) & 2.3(2)  & 0.13(2)  & 0.67(7) & 2.9(2)  & 0.12(3)  & 0.78(8) & 3.7(3)  & 0.11(4) \\ 
150 & 0.60(7) & 2.3(2) & 0.13(2)  & 0.62(7) & 2.4(2)  & 0.13(2)  & 0.67(7) & 2.8(2)  & 0.12(2)  & 0.77(7) & 3.4(2)  & 0.12(2)  & 0.90(8) & 4.3(2)  & 0.12(2) \\ 
155 & 0.68(7) & 2.7(2) & 0.13(3)  & 0.70(7) & 2.9(2)  & 0.13(3)  & 0.77(7) & 3.3(2)  & 0.13(3)  & 0.88(7) & 3.9(2)  & 0.13(3)  & 1.03(8) & 4.8(2)  & 0.12(3) \\ 
160 & 0.77(7) & 3.1(2) & 0.15(3)  & 0.79(7) & 3.2(2)  & 0.14(3)  & 0.86(7) & 3.6(2)  & 0.14(3)  & 0.99(7) & 4.2(2)  & 0.14(3)  & 1.16(8) & 5.1(2)  & 0.14(3) \\ 
165 & 0.87(7) & 3.4(2) & 0.16(2)  & 0.90(7) & 3.5(2)  & 0.16(2)  & 0.97(7) & 3.8(2)  & 0.16(2)  & 1.11(7) & 4.4(2)  & 0.16(2)  & 1.29(7) & 5.2(2)  & 0.16(3) \\ 
170 & 0.97(7) & 3.6(2) & 0.17(2)  & 1.00(7) & 3.7(2)  & 0.17(2)  & 1.09(7) & 4.0(2)  & 0.17(2)  & 1.23(7) & 4.5(2)  & 0.18(2)  & 1.42(7) & 5.2(2)  & 0.18(2) \\ 
175 & 1.08(6) & 3.8(2) & 0.18(2)  & 1.11(6) & 3.9(2)  & 0.18(2)  & 1.20(6) & 4.1(2)  & 0.18(2)  & 1.34(6) & 4.6(2)  & 0.19(2)  & 1.55(6) & 5.2(2)  & 0.19(2) \\ 
180 & 1.19(6) & 3.9(1) & 0.19(2)  & 1.22(6) & 4.0(1)  & 0.19(2)  & 1.31(6) & 4.2(1)  & 0.19(2)  & 1.46(6) & 4.6(1)  & 0.20(2)  & 1.66(6) & 5.1(1)  & 0.20(2) \\ 
185 & 1.30(5) & 4.0(1) & 0.20(2)  & 1.33(5) & 4.1(1)  & 0.20(2)  & 1.42(5) & 4.3(1)  & 0.20(2)  & 1.57(6) & 4.6(1)  & 0.21(2)  & 1.78(6) & 5.0(1)  & 0.21(2) \\ 
190 & 1.40(6) & 4.0(1) & 0.21(2)  & 1.43(6) & 4.1(1)  & 0.21(2)  & 1.52(6) & 4.3(1)  & 0.21(2)  & 1.67(6) & 4.5(1)  & 0.22(2)  & 1.88(6) & 4.9(1)  & 0.22(2) \\ 
200 & 1.61(6) & 4.0(2) & 0.23(1)  & 1.64(6) & 4.1(2)  & 0.22(1)  & 1.73(6) & 4.2(2)  & 0.22(1)  & 1.87(6) & 4.4(2)  & 0.23(1)  & 2.08(6) & 4.7(2)  & 0.24(1) \\ 
220 & 1.98(7) & 3.8(2) & 0.25(1)  & 2.01(7) & 3.8(2)  & 0.24(1)  & 2.09(7) & 3.9(2)  & 0.24(1)  & 2.22(7) & 4.0(2)  & 0.25(1)  & 2.41(7) & 4.1(2)  & 0.25(1) \\ 
240 & 2.29(7) & 3.4(1) & 0.27(2)  & 2.32(7) & 3.4(1)  & 0.26(2)  & 2.39(7) & 3.4(1)  & 0.26(2)  & 2.51(7) & 3.5(1)  & 0.27(2)  & 2.67(7) & 3.6(1)  & 0.27(2) \\ 
260 & 2.55(7) & 3.0(1) & 0.28(2)  & 2.58(7) & 3.0(1)  & 0.27(2)  & 2.64(7) & 3.1(1)  & 0.28(2)  & 2.74(7) & 3.1(1)  & 0.28(2)  & 2.89(7) & 3.1(1)  & 0.28(2) \\ 
280 & 2.77(7) & 2.8(2) & 0.28(2)  & 2.79(7) & 2.8(2)  & 0.28(2)  & 2.84(7) & 2.8(2)  & 0.28(2)  & 2.93(7) & 2.8(2)  & 0.28(2)  & 3.06(7) & 2.8(2)  & 0.28(2) \\ 
300 & 2.95(7) & 2.6(2) & 0.29(2)  & 2.96(7) & 2.6(2)  & 0.28(2)  & 3.01(7) & 2.6(2)  & 0.28(2)  & 3.09(7) & 2.6(2)  & 0.29(2)  & 3.21(7) & 2.6(2)  & 0.29(2) \\ 
320 & 3.10(7) & 2.3(2) & 0.30(2)  & 3.12(7) & 2.3(2)  & 0.29(2)  & 3.16(7) & 2.3(2)  & 0.29(2)  & 3.23(7) & 2.3(2)  & 0.29(2)  & 3.33(7) & 2.4(2)  & 0.29(2) \\ 
340 & 3.24(8) & 2.1(2) & 0.30(3)  & 3.25(8) & 2.1(2)  & 0.30(3)  & 3.29(8) & 2.1(2)  & 0.30(3)  & 3.35(8) & 2.1(2)  & 0.30(3)  & 3.44(8) & 2.1(2)  & 0.30(3) \\ 
360 & 3.35(8) & 1.9(2) & 0.31(3)  & 3.36(8) & 1.9(2)  & 0.31(3)  & 3.40(8) & 1.9(2)  & 0.31(3)  & 3.45(8) & 1.9(2)  & 0.31(3)  & 3.53(8) & 1.9(2)  & 0.31(3) \\ 
380 & 3.44(8) & 1.7(1) & 0.31(3)  & 3.45(8) & 1.7(1)  & 0.31(3)  & 3.49(8) & 1.7(1)  & 0.31(3)  & 3.54(8) & 1.7(1)  & 0.31(3)  & 3.61(8) & 1.7(1)  & 0.31(3) \\ 
400 & 3.52(8) & 1.5(1) & 0.31(3)  & 3.53(8) & 1.5(1)  & 0.31(3)  & 3.56(8) & 1.5(1)  & 0.31(3)  & 3.61(8) & 1.5(1)  & 0.31(3)  & 3.67(8) & 1.5(1)  & 0.31(3) \\ 
\hline
\end{tabular}